\documentclass[11pt]{article}
\usepackage{amssymb}

\usepackage{graphicx}
\usepackage{amsmath}
\usepackage{makeidx}
\usepackage{indentfirst}

\newcounter{resultnum}[section]

\newcounter{conclusionnum}[section]

\newcounter{conditionnum}[section]

\newcounter{conjecturenum}[section]

\newcounter{examplenum}[section]

\newcounter{exercisenum}[section]

\newcounter{lemmanum}[section]

\newcounter{notationnum}[section]

\newcounter{theoremnum}[section]

\newcounter{definitionnum}[section]

\newcounter{corollarynum}[section]

\newcounter{remarknum}[section]

\newcounter{propositionnum}[section]

\newcounter{acknowledgementnum}[section]

\newcounter{algorithmnum}[section]

\newcounter{axiomnum}[section]

\newcounter{casenum}[section]

\newcounter{claimnum}[section]

\newcounter{summarynum}[section]

\newcounter{problemnum}[section]

\begin{document}

\title{Fractional Dynamics from Einstein Gravity,\\
General Solutions, and Black Holes}
\date{April 5, 2010}
\author{\textbf{Sergiu I. Vacaru} \thanks{%
sergiu.vacaru@uaic.ro, Sergiu.Vacaru@gmail.com} \\
\textsl{\small University "Al. I. Cuza" Ia\c si, Science Department,} \\
\textsl{\small 54 Lascar Catargi street, Ia\c si, Romania, 700107 }}
\maketitle

\begin{abstract}
We study the fractional gravity for spacetimes with non--integer dimensions. Our constructions are based on a geometric formalism with the fractional Caputo derivative and integral calculus adapted to nonolonomic distributions. This allows us to define a fractional spacetime geometry with
fundamental geometric/physical objects and a generalized tensor calculus all being similar to respective integer dimension constructions. Such models of fractional gravity mimic the Einstein gravity theory and various Lagrange--Finsler and Hamilton--Cartan generalizations in nonholonomic
variables. The approach suggests a number of new implications for gravity and matter field theories with singular, stochastic, kinetic, fractal, memory etc processes. We prove that the fractional gravitational field equations can be integrated in very general forms following the anholonomic
deformation method for constructing exact solutions. Finally, we study some examples of fractional black hole solutions, fractional ellipsoid gravitational configurations and imbedding of such objects in fractional solitonic backgrounds.

\vskip0.3cm


\vskip3pt


\end{abstract}

\tableofcontents


\section{Introduction}

Modern classical and quantum gravity models are constructed geometrically to
be extremely complex with black holes, singularities and horizons,
nontrivial topology, stochastic processes and fractals, irregular sets,
possible extra dimensions etc. Such configurations are defined as exact
and/or approximate solutions of Einstein equations in general relativity and
modifications.

There is a recent interest in fractional analysis and non--integer dimension
geometries which resulted in an increasing number of publications during
last decade. Calculus with derivatives and integrals of fractional order is
applied in various directions in physics, for instance, in Lagrange and
Hamilton mechanics and dynamical models \cite%
{riewe97,balmus05,muslbal05,taras06,bal06}, fractal and chaotic dynamics %
\cite{carp,hilfer,taras05,zasl1,zasl2}, quantum mechanics \cite%
{lask1,lask2,nab04}, kinetic theories \cite{zasl1,zasl3,taraszasl}, plasma
physics \cite{carrer,taraspl} and astrophysics and cosmology \cite%
{tarasgr,nab1,nab2,roberts}, and many additional applications in physics and
other sciences \cite{west,boleantu,magin,sommacal}.

The cornerstone questions to be solved in constructing classical and quantum
gravity theories are positively connected to the meaning of spacetime
dimension, extra dimension and/or fractional dimensions physical effects,
possible contributions from (non) holonomic and/or commutative/
noncommutative variables, of fractal dimension, nonlocal field theories etc.
With respect to dark energy and dark matter models, various attempts to
construct the quantum gravity theory, to determine the status of
singularities in fundamental theories etc, it is thus natural to pursue
alternative concepts of differential and integral calculus and consider new
ideas and models of spacetime geometry.

The fractional dimension paradigm is at present extended to new ideas on
gauge invariance in fractional field theories \cite{herrmann}, being
constructed solutions of fractional Dirac equation \cite{raspini2,mab}, with
corresponding formalisms for generalized Clifford algebra \cite{zavada}. A
formulation of Noether's theorem for fractional classical fields \cite{musln}
and, recently, a model of fractional gravity based on Riemann--Liouville
concept of non--integer dimensional derivative was proposed in \cite{munk}.

Let us provide a simple motivation why knew mathematical tools from
non--integer calculus may be very useful in solving fundamental theories in
physics. An exact solution of Einstein equations with singularity for $%
r\rightarrow r_{0}$ (for instance, inducing a singular term in the Riemann
curvature tensor) modelled by a function
\begin{equation}
f(r)\sim \ ^{1}f(r_{0})(r-r_{0})^{-\ ^{1}s}+\ ^{2}f(r_{0})(r-r_{0})^{\
^{2}s}+...,  \label{sfunct}
\end{equation}%
for some real values $\ ^{1}f(r_{0})$ and $\ ^{2}f(r_{0})$ and positive
integers $\ ^{1}s$ and $\ ^{2}s.$ The singular properties of such a function
may have a complete different geometric and classical/quantum physical
meaning if instead of the standard partial derivative $\partial /\partial r$
we work with the left Caputo fractional derivative $\ _{\ _{1}r}\overset{%
\alpha }{\underline{\partial }}_{r}f(r)$ (when usual partial derivatives are
generalized through a more complex integral--derivative relation, see below
formula (\ref{lfcd})).\footnote{%
We follow the \ system of notations with left abstract labels explained in
Refs. \cite{vfrrf,vrflg}, see also next section.} Such functions positively
result in various new physical implications if there are developed models
with inhomogeneous, fractal, stochastic, kinetic etc processes, in classical
and quantum theories.

The advantages of fractional derivatives and integrals become already
apparent in modern mechanics and for describing thermodynamical and
electrical properties of real complex media (as we emphasized in the above
references). We suppose that for non--integer dimensions, new conceptual
features of gravitational nonlinear interactions can be analyzed.
Fractional--order models are more appropriate than the integer--order ones
because they provide an excellent instrument for describing processes with
memory, branching and hereditary properties which in the classical (pseudo)
Riemannian approach is, in fact, neglected (even such solutions can be
encoded into the structure of certain quantum/stochastic modifications of
Einstein equations).

In our recent work \cite{vfrrf}, we provided some theoretical arguments that
fractional spacetime theories should receive more attention in modern
geometric analysis, evolution theories and gravity. \ Such a conclusion was
derived from the facts that the Ricci flow theory can be naturally
generalized for non--integer dimensions, with a new type of statistical
geometric analogy, if a corresponding nonholonomic fractional
differential--integral calculus is applied. Such a scenario may transform
(pseudo) Riemannian metrics into certain fractional ones, or inversely. This
involve a number of consequences in classical and quantum theories of
fundamental field interactions and evolution models. Here we also remember
that the theory of fractional calculus with derivatives and integrals of
non--integer order goes back to Leibniz, Louville, Grunwald, Letnikov and
Riemann \cite{oldham,podlubny,kilbas,nishimoto} (The first example of
derivative of order $\alpha =1/2$ was considered by Leibnitz in 1695, see
historical remarks in \cite{ross}).

One might hope that extending a fundamental physical theory like general
gravity to fractional dimensions will solve the most troublesome problems of
Planckian physics, quantum non-renormalizable theories, new types of
symmetries etc. Such extensions should be constructive ones allowing exact
solutions and possible simple geometric and physical interpretations. In
this paper, we shall apply the fractional geometric formalism developed in
Ref. \cite{vfrrf} to formulate and study a version of fractional Einstein
type gravity theory. Certain constructions with the fractional Ricci tensor,
and generalizations to osculator bundles (higher order fractional mechanical
models etc) and non--integer dimension Einstein tensor were provided in
Refs. \cite{albu,munk}. It is a very cumbersome task, both technically and
theoretically, to solve field and evolution fractional equations in exact
form following those approaches with the Riemann--Liouville (RL) fractional
derivative.

In general, there are various possibilities to derive fractional
differential--integral structures and the question on which derivative would
be more suitable for formulating engineering and scientific problems remain
an open issue. Here we note that integrals on net fractals can be
approximated by the left--sided RL fractional integrals of functions, see %
\cite{rqlw}. Such a fractal and non--integer calculus formalism is applied
for certain attempts to construct the quantum field theory, gravity and
cosmology in fractal universes and related diffusion processes in string
theory, see \cite{calcagni1,calcagni2} and references therein. Nevertheless,
we shall follow an approach based on Caputo fractional derivatives allowing,
for instance, to get nonhomogeneous cosmological solutions generalizing the
off--diagonal metrics from \cite{vcosms}.

This paper has three purposes: The first one is to formulate a fractional
generalization of the Einstein gravitational field equations with
nonholonomic local frames and deformations of connections induced by the
Caputo left derivative. The second one is to prove that such nonlinear
systems of partial derivative and integral equations can be solved in very
general form using our anholonomic deformation method of constructing exact
solutions in gravity and Ricci flow theories, see reviews of results and
applications in \cite{vncg,vsgg,ijgmmp,vrflg,vexsol,vcg}. The third purpose,
is to study certain applications of fractional calculus in black hole
physics: we construct in explicit form such fractional black hole solutions
and study their imbedding in fractional solitonic backgrounds.

The article is structured as follows. In section \ref{sfc} we provide an
introduction into fractional Caputo type calculus on nonholonomic manifolds.%
\footnote{%
In "integer" differential geometry, there are used also equivalent terms
like anholonomic and non--integrable manifolds. A nonholonomic manifold of
integer dimension is defined by a pair $(\mathbf{V},\mathcal{N})$, where $%
\mathbf{V}$ is a manifold and $\mathcal{N}$ is a nonintegrable distribution
on $\mathbf{V},$ see the next section for a generalization to non--integer
dimensions.} The Einstein equations are generalized on fractional
nonholonomic manifold in section \ref{seefnm}. Then, in section \ref{sesfee}
we show that in our approach the fractional field equations for gravity can
be integrated in very general forms. We provide some applications in modern
gravity by considering black hole solutions in factional spacetimes, see
section \ref{sbhfs}. Discussion of results and future perspectives are
presented in concluding section \ref{sconcl}.

\section{Fractional Derivatives, Integrals and Forms}

\label{sfc}

We provide an introduction to fractional differential and integral calculus
on ''flat'' spaces following \cite{taras08} and then extend the approach as
fractional generalizations of (nonholonomic) Einstein spacetimes. Relevant
details and references can be found also in a partner paper \cite{vfrrf}
(see there and Appendix containing a summary of non--integer calculus and
methods).

\subsection{Fractional Caputo derivatives}

There are different approaches to fractional derivative and integral
calculus when there are used different types of fractional derivatives. We
elaborate and follow a method with nonholonomic distributions when the
geometric constructions are most closed to ''integer'' calculus.

Let us consider that $f(x)$ is a derivable function $f:[\ _{1}x,\
_{2}x]\rightarrow \mathbb{R},$ for $\mathbb{R\ni }\alpha >0,$ and denote the
derivative on $x$ as $\partial _{x}=\partial /\partial x.$ The fractional
Caputo derivatives are determined respectively by
\begin{eqnarray}
&&\mbox{left:\ } \ _{\ _{1}x}\overset{\alpha }{\underline{\partial }}%
_{x}f(x):=\frac{1}{\Gamma (s-\alpha )}\int\limits_{\ \ _{1}x}^{x}(x-\
x^{\prime })^{s-\alpha -1}\left( \frac{\partial }{\partial x^{\prime }}%
\right) ^{s}f(x^{\prime })dx^{\prime };  \label{lfcd} \\
&&\mbox{right:\ } \ _{\ x}\overset{\alpha }{\underline{\partial }}_{\
_{2}x}f(x):=\frac{1}{\Gamma (s-\alpha )}\int\limits_{x}^{\ _{2}x}(x^{\prime
}-x)^{s-\alpha -1}\left( -\frac{\partial }{\partial x^{\prime }}\right)
^{s}f(x^{\prime })dx^{\prime }\ ,  \notag
\end{eqnarray}%
where we underline the partial derivative symbol, $\underline{\partial },$
in order to distinguish the Caputo operators from the RL ones with usual $%
\partial .$\footnote{%
The left Riemann--Liouville (RL) derivative is
\begin{equation*}
\ \ _{\ _{1}x}\overset{\alpha }{\partial }_{x}f(x):=\frac{1}{\Gamma
(s-\alpha )}\left( \frac{\partial }{\partial x}\right) ^{s}\int\limits_{\ \
_{1}x}^{x}(x-\ x^{\prime })^{s-\alpha -1}f(x^{\prime })dx^{\prime },
\end{equation*}%
where $\Gamma $ is the Euler's gamma function. The left fractional Liouville
derivative of order $\alpha ,$ where $\ s-1<\alpha <s,$ with respect to
coordinate $x$ is $\ \overset{\alpha }{\partial }_{x}f(x):=\lim_{_{1}x%
\rightarrow -\infty }\ \ _{\ _{1}x}\overset{\alpha }{\partial }_{x}f(x).$%
\par
The right RL derivative is $\ _{\ x}\overset{\alpha }{\partial }_{\
_{2}x}f(x):=\frac{1}{\Gamma (s-\alpha )}\left( -\frac{\partial }{\partial x}%
\right) ^{s}\int\limits_{x}^{\ _{2}x}(x^{\prime }-x)^{s-\alpha
-1}f(x^{\prime })dx^{\prime }\ .$ The right fractional Liouville derivative
is $\ _{\ x}\overset{\alpha }{\partial }f(x^{k}):=\lim_{_{2}x\rightarrow
\infty }\ \ _{\ x}\overset{\alpha }{\partial }_{\ _{2}x}f(x).$%
\par
The fractional RL derivative of a constant $C$ is not zero but, for
instance, $\ _{\ _{1}x}\overset{\alpha }{\partial }_{x}C=C\frac{(x-\
_{1}x)^{-\alpha }}{\Gamma (1-\alpha )}.$ Integro--differential constructions
based only on such derivatives seem to be very cumbersome and has a number
of properties which are very different from similar ones for the integer
calculus.} A very important property is that for a constant $C,$ for
instance, $_{\ _{1}x}\overset{\alpha }{\underline{\partial }}_{x}C=0.$

\subsubsection{Fractional integral}

We denote by $L_{z}(\ _{1}x,\ _{2}x)$ the set of those Lesbegue measurable
functions $f$ on $[\ _{1}x,\ _{2}x]$ for which $||f||_{z}=(\int%
\limits_{_{1}x}^{_{2}x}|f(x)|^{z}dx)^{1/z}<\infty $ and write $C^{z}[\
_{1}x,\ _{2}x]$ for a space of functions which are $z$ times continuously
differentiable on this interval.

For any real--valued function $f(x)$ defined on a closed interval $[\
_{1}x,\ _{2}x],$ there is a function $F(x)=_{\ _{1}x}\overset{\alpha }{I}%
_{x}\ f(x)$ defined by the fractional Riemann--Liouville integral $\ _{\
_{1}x}\overset{\alpha }{I}_{x}f(x):=\frac{1}{\Gamma (\alpha )}%
\int\limits_{_{1}x}^{x}(x-x^{\prime })^{\alpha -1}f(x^{\prime })dx^{\prime
}, $ when the function $f(x)=\ _{\ _{1}x}\overset{\alpha }{\underline{%
\partial }}_{x}F(x),$ for all $x\in \lbrack \ _{1}x,\ _{2}x],$\footnote{%
this follows from the fundamental theorem of fractional calculus \cite%
{taras08}} satisfies the conditions
\begin{eqnarray*}
\ _{\ _{1}x}\overset{\alpha }{\underline{\partial }}_{x}\left( _{\ _{1}x}%
\overset{\alpha }{I}_{x}f(x)\right) &=&f(x),\ \alpha >0, \\
_{\ _{1}x}\overset{\alpha }{I}_{x}\left( \ _{\ _{1}x}\overset{\alpha }{%
\underline{\partial }}_{x}F(x)\right) &=&F(x)-F(\ _{1}x),\ 0<\alpha <1.
\end{eqnarray*}

A fractional volume integral is a triple fractional integral within a region
$X\subset \mathbb{R}^{3},$ for instance, of a scalar field $f(x^{k}),$%
\begin{equation*}
\overset{\alpha }{I}(f)=\overset{\alpha }{I}[x^{k}]f(x^{k})=\overset{\alpha }%
{I}[x^{1}]\ \overset{\alpha }{I}[x^{2}]\ \overset{\alpha }{I}[x^{3}]f(x^{k}).
\end{equation*}%
For $\alpha =1$ and $f(x,y,z),$ $\overset{\alpha }{I}(f)=\iiint%
\limits_{X}dVf(x,y,z)= \iiint\limits_{X}dxdydz\ f(x,y,z).$

\subsubsection{Fractional differential forms}

An exterior fractional differential can be defined through the fractional
Caputo derivatives which is self--consistent with the definition of the
fractional integral considered above. We write the fractional absolute
differential $\overset{\alpha }{d}$ in the form
\begin{equation*}
\overset{\alpha }{d}:=(dx^{j})^{\alpha }\ \ _{\ 0}\overset{\alpha }{%
\underline{\partial }}_{j},\mbox{ where }\ \overset{\alpha }{d}%
x^{j}=(dx^{j})^{\alpha }\frac{(x^{j})^{1-\alpha }}{\Gamma (2-\alpha )},
\end{equation*}%
where we consider $\ _{1}x^{i}=0.$\footnote{%
For the ''integer'' calculus, we use as local coordinate co-bases/--frames
the differentials $dx^{j}=(dx^{j})^{\alpha =1}.$ For $0<\alpha <1$ we have $%
dx=(dx)^{1-\alpha }(dx)^{\alpha }.$ $\ $The ''fractional'' symbol $%
(dx^{j})^{\alpha },$ related to $\overset{\alpha }{d}x^{j},$ is used instead
of $dx^{i}$ for elaborating a co--vector/differential form calculus, see
below the formula (\ref{frdif}).}

An exterior fractional differential is defined
\begin{equation*}
\overset{\alpha }{d}=\sum\limits_{j=1}^{n}\Gamma (2-\alpha )(x^{j})^{\alpha
-1}\ \overset{\alpha }{d}x^{j}\ \ _{\ 0}\overset{\alpha }{\underline{%
\partial }}_{j}.
\end{equation*}%
Differentials are dual to partial derivatives, and derivation is inverse to
integration. For a fractional calculus, \ the concept of ''dual'' and
''inverse'' have a more sophisticate relation to ''integration'' and, in
result, there is a more complex relation between forms and vectors.

The fractional integration for differential forms on an interval\newline
$L=[\ _{1}x,\ _{2}x]$ is defined
\begin{equation}
\ _{L}\overset{\alpha }{I}[x]\ \ _{\ _{1}x}\overset{\alpha }{d}_{x}f(x)=f(\
_{2}x)-f(\ _{1}x),  \label{aux01}
\end{equation}%
i.e. the fractional differential of a function $\ f(x)$ is $_{\ _{1}x}%
\overset{\alpha }{d}_{x}f(x)=[...],$ when
\begin{equation*}
\int\limits_{_{1}x}^{_{2}x}\frac{(dx)^{1-\alpha }}{\Gamma (\alpha )(\
_{2}x-x)^{1-\alpha }}[(dx^{\prime })^{\alpha }\ _{\ _{1}x}\overset{\alpha }{%
\underline{\partial }}_{x^{\prime \prime }}f(x^{\prime \prime })] =f(x)-f(\
_{1}x).
\end{equation*}

The exact fractional differential 0--form is a fractional differential of
the function
\begin{equation*}
\ _{\ _{1}x}\overset{\alpha }{d}_{x}f(x):=(dx)^{\alpha }\ \ _{\ _{1}x}%
\overset{\alpha }{\underline{\partial }}_{x^{\prime }}f(x^{\prime }),
\end{equation*}%
when the equation (\ref{aux01}) is considered as the fractional
generalization of the integral for a differential 1--form. So, the formula
for the fractional exterior derivative can be written
\begin{equation}
\ \ _{\ _{1}x}\overset{\alpha }{d}_{x}:=(dx^{i})^{\alpha }\ \ _{\ _{1}x}%
\overset{\alpha }{\underline{\partial }}_{i}.  \label{feder}
\end{equation}%
For instance, for the fractional differential 1--form $\ \overset{\alpha }{%
\omega }$ with coefficients $\{\omega _{i}(x^{k})\}$ is
\begin{equation}
\ \overset{\alpha }{\omega }=(dx^{i})^{\alpha }\ \omega _{i}(x^{k})
\label{fr1f}
\end{equation}%
and the exterior fractional derivatives of such a fractional 1--form $\
\overset{\alpha }{\omega }$ gives a fractional 2--form, $\ _{\ _{1}x}\overset%
{\alpha }{d}_{x}(\ \overset{\alpha }{\omega })=(dx^{i})^{\alpha }\wedge
(dx^{j})^{\alpha }\ _{\ _{1}x}\overset{\alpha }{\underline{\partial }}_{j}\
\omega _{i}(x^{k}).$ Such a rule \cite{kilbas} follows from the property
that for any type fractional derivative $\ \overset{\alpha }{\partial }_{x\
},$ we have
\begin{equation*}
\ \overset{\alpha }{\partial }_{x\ }(\ ^{1}f\ \
^{2}f)=\sum\limits_{k=0}^{\infty }\left(
\begin{array}{c}
\alpha \\
k%
\end{array}%
\right) \left( \overset{\alpha -k}{\partial }_{x\ }\ ^{1}f\right) \ \
\overset{\alpha =k}{\partial }_{x\ }\left( \ ^{2}f\right) ,
\end{equation*}%
when for integer $k,$
\begin{equation*}
\left(
\begin{array}{c}
\alpha \\
k%
\end{array}%
\right) =\frac{(-1)^{k-1}\alpha \Gamma (k-\alpha )}{\Gamma (1-\alpha )\Gamma
(k+1)}\mbox{\ and \ } \ \overset{k}{\partial }_{x\ }\ (dx)^{\alpha }=0,k\geq
1.
\end{equation*}

In fractional derivative calculus, there are used also the properties:\
\begin{equation*}
\ _{_{1}x^{\prime }}\overset{\alpha }{\partial }_{x^{\prime }\ }(x^{\prime
}-\ _{1}x^{\prime })^{\beta }=\frac{\Gamma (\beta +1)}{\Gamma (\beta
+1-\alpha )}(x-\ _{1}x)^{\beta -\alpha },
\end{equation*}
where $n-1<\alpha <n$ and $\beta >n,$ and $\ _{_{1}x^{\prime }}\overset{%
\alpha }{\partial }_{x^{\prime }\ }(x^{\prime }-\ _{1}x^{\prime })^{k}=0$
for $k=0,1,2,...,n-1.$ We obtain $\ $%
\begin{equation*}
_{\ _{1}x}\overset{\alpha }{d}_{x}(x-\ _{1}x)^{\alpha }=(dx)^{\alpha }\ _{\
_{1}x}\overset{\alpha }{\underline{\partial }}_{i^{\prime }}x^{i^{\prime
}}=(dx)^{\alpha }\Gamma (\alpha +1),
\end{equation*}
i.e.
\begin{equation}
(dx)^{\alpha }=\frac{1}{\Gamma (\alpha +1)}\ _{\ _{1}x}\overset{\alpha }{d}%
_{x}(x-\ _{1}x)^{\alpha }.  \label{frdif}
\end{equation}%
We might write the fractional exterior derivative (\ref{feder}) in the form
\begin{equation*}
\ \ _{\ _{1}x}\overset{\alpha }{d}_{x}:=\frac{1}{\Gamma (\alpha +1)}\ \ _{\
_{1}x}\overset{\alpha }{d}_{x}(x^{i}-\ _{1}x^{i})^{\alpha }\ _{\ _{1}x}%
\overset{\alpha }{\underline{\partial }}_{i}
\end{equation*}%
and the \ fractional differential 1--form (\ref{fr1f}) as%
\begin{equation*}
\ \overset{\alpha }{\omega }=\frac{1}{\Gamma (\alpha +1)}\ _{\ _{1}x}%
\overset{\alpha }{d}_{x}(x^{i}-\ _{1}x^{i})^{\alpha }\ F_{i}(x).
\end{equation*}

Having a well defined exterior calculus of fractional differential forms on
flat spaces $\mathbb{R}^{n},$ we can generalize the constructions for a real
manifold \ $M,\dim M=n.$\

\subsection{Fractional manifolds and tangent bundles}

A real manifold $M,$ with integer dimension $\dim M=n,$ can be endowed on charts of a covering atlas with
a fractional derivative--integral structure of Caputo type as we explained
above. In brief, such a space (of necessary smooth class)$\ \overset{\alpha }%
{\underline{M}}$ \ will be called a fractional manifold.

Let us explain how the concept of tangent bundle can be developed for
fractional dimensions. A tangent bundle $TM$ over a manifold $M$ of
integer dimension is canonically defined by its local integer differential
structure $\partial _{i}$. A fractional generalization can be obtained if
instead of $\partial _{i}$ we consider the left Caputo derivatives $\ _{\
_{1}x^{i}}\overset{\alpha }{\underline{\partial }}_{i}$ of type (\ref{lfcd}%
), for every local coordinate $x^{i}$ on a local cart $X$ on $M.$ A
fractional tangent bundle $\overset{\alpha }{\underline{T}}M$ \ for $\alpha
\in (0,1)$ (the symbol $T$ is underlined in order to emphasize that we shall
associate the approach to a fractional Caputo derivative), see more details
in \cite{vfrrf}.\footnote{%
We to not follow alternative constructions with RL fractional derivative %
\cite{albu,munk} because in that direction it is not clear how to prove
integrability of the RL--fractional Einstein equations.} For simplicity, we
shall write both for integer and fractional tangent bundles the local
coordinates in the form $u^{\beta }=(x^{j},y^{j}).$

On $\overset{\alpha }{\underline{T}}M,$ an arbitrary fractional frame basis
\begin{equation}
\overset{\alpha }{\underline{e}}_{\beta }=e_{\ \beta }^{\beta ^{\prime
}}(u^{\beta })\overset{\alpha }{\underline{\partial }}_{\beta ^{\prime }}
\label{flcfb}
\end{equation}%
is connected via a vierlbein transform $e_{\ \beta }^{\beta ^{\prime
}}(u^{\beta })$ with a fractional local coordinate basis
\begin{equation}
\overset{\alpha }{\underline{\partial }}_{\beta ^{\prime }}=\left( \overset{%
\alpha }{\underline{\partial }}_{j^{\prime }}=_{\ _{1}x^{j^{\prime }}}%
\overset{\alpha }{\underline{\partial }}_{j^{\prime }},\overset{\alpha }{%
\underline{\partial }}_{b^{\prime }}=_{\ _{1}y^{b^{\prime }}}\overset{\alpha
}{\underline{\partial }}_{b^{\prime }}\right)  \label{frlcb}
\end{equation}%
when $j^{\prime }=1,2,...,n$ and $b^{\prime }=n+1,n+2,...,n+n.$ \ There are
also fractional co--bases which are dual to (\ref{flcfb}),
\begin{equation}
\overset{\alpha }{\underline{e}}^{\ \beta }=e_{\beta ^{\prime }\ }^{\ \beta
}(u^{\beta })\overset{\alpha }{d}u^{\beta ^{\prime }},  \label{flcfdb}
\end{equation}%
where the fractional local coordinate co--basis
\begin{equation}
\ _{\ }\overset{\alpha }{d}u^{\beta ^{\prime }}=\left( (dx^{i^{\prime
}})^{\alpha },(dy^{a^{\prime }})^{\alpha }\right) ,  \label{frlccb}
\end{equation}%
when the h-- and v--components, $(dx^{i^{\prime }})^{\alpha }$ and $%
(dy^{a^{\prime }})^{\alpha }$ are of type (\ref{fr1f}). For integer values,
a matrix $e_{\ \beta }^{\beta ^{\prime }}$ is inverse to $e_{\beta
^{\prime}\ }^{\ \beta }.$

Similarly to $\overset{\alpha }{\underline{T}}M,$ we can define a fractional
vector bundle $\overset{\alpha}{\underline{E}}$ on $M,$ when the fiber
indices of bases run values $a^{\prime },b^{\prime },...=n+1,n+2,...,n+m.$

\section{Einstein Equations on Fractional Manifolds}

\label{seefnm} \ We provide a fractional generalization of the Einstein
gravity of arbitrary dimensions via nonholonomic variables in such a form
when we shall be able to construct exact solutions of the gravitational
field equations.

\subsection{Introduction to the geometry of fractional manifolds}

Let us consider a ''prime'' (pseudo) Riemannian manifold $\mathbf{V}$ is of
integer dimension $\dim $ $\mathbf{V}=n+m,n\geq 2,m\geq 1.$\ Its fractional
extension is modelled as a fractional nonholonomic manifold $\overset{\alpha
}{\mathbf{V}}$ defined by a quadruple $(\mathbf{V},\overset{\alpha }{%
\mathcal{N}},\overset{\alpha }{\mathbf{d}},\overset{\alpha }{\mathbf{I}}),$
where $\overset{\alpha }{\mathcal{N}}$ \ (see below formula (\ref{whit})) is
a nonholonomic distribution defining a nonlinear connection structure, the
fractional differential structure $\overset{\alpha }{\mathbf{d}}$ is given
by (\ref{flcfb}) and (\ref{flcfdb}) and the non--integer integral structure $%
\overset{\alpha }{\mathbf{I}}$ is computed following the by rules of type (%
\ref{aux01}).

\subsubsection{Nonholonomic fractional distributions and frames}

A nonlinear connection (N--connection) $\overset{\alpha }{\mathbf{N}}$ \ for
$\overset{\alpha }{\mathbf{V}}$ is defined by a nonholonomic distribution
(Whitney sum) with conventional h-- and v--subspaces, $\underline{h}\overset{%
\alpha }{\mathbf{V}}$ and $\underline{v}\overset{\alpha }{\mathbf{V}},$%
\begin{equation}
\overset{\alpha }{\underline{T}}\overset{\alpha }{\mathbf{V}}=\underline{h}%
\overset{\alpha }{\mathbf{V}}\mathbf{\oplus }\underline{v}\overset{\alpha }{%
\mathbf{V}}.  \label{whit}
\end{equation}

Nonholonomic manifolds with a $\overset{\alpha }{\mathbf{N}}$ are called, in
brief, N--anholonomic fractional manifolds. Locally, a fractional
N--connection is defined by its coefficients, $\overset{\alpha }{\mathbf{N}}%
\mathbf{=}\{\ ^{\alpha }N_{i}^{a}\},$ when
\begin{equation}
\overset{\alpha }{\mathbf{N}}\mathbf{=}\ ^{\alpha
}N_{i}^{a}(u)(dx^{i})^{\alpha }\otimes \overset{\alpha }{\underline{\partial
}}_{a}.  \label{fnccoef}
\end{equation}%
For a N--connection $\overset{\alpha }{\mathbf{N}},$ we can always a class
of fractional (co) frames (N--adapted) linearly depending on $\ ^{\alpha
}N_{i}^{a}),$
\begin{eqnarray}
\ ^{\alpha }\mathbf{e}_{\beta }&=&\left[ \ ^{\alpha }\mathbf{e}_{j}=\overset{%
\alpha }{\underline{\partial }}_{j}-\ ^{\alpha }N_{j}^{a}\overset{\alpha }{%
\underline{\partial }}_{a},\ ^{\alpha }e_{b}=\overset{\alpha }{\underline{%
\partial }}_{b}\right],  \label{dder} \\
\ ^{\alpha }\mathbf{e}^{\beta }&=&[\ ^{\alpha }e^{j}=(dx^{j})^{\alpha },\
^{\alpha }\mathbf{e}^{b}=(dy^{b})^{\alpha }+\ ^{\alpha
}N_{k}^{b}(dx^{k})^{\alpha }].  \label{ddif}
\end{eqnarray}%
The nontrivial nonholonomy coefficients are computed $\ ^{\alpha }W_{ib}^{a}=%
\overset{\alpha }{\underline{\partial }}_{b}\ ^{\alpha }N_{i}^{a}$ and $\
^{\alpha }W_{ij}^{a}=\ ^{\alpha }\Omega _{ji}^{a}=\ ^{\alpha }\mathbf{e}%
_{i}\ ^{\alpha }N_{j}^{a}-\ ^{\alpha }\mathbf{e}_{j}\ ^{\alpha }N_{i}^{a}$
(where $\ ^{\alpha }\Omega _{ji}^{a}$ are the coefficients of the
N--connection curvature) for
\begin{equation*}
\left[ \ ^{\alpha }\mathbf{e}_{\alpha },\ ^{\alpha }\mathbf{e}_{\beta }%
\right] =\ ^{\alpha }\mathbf{e}_{\alpha }\ ^{\alpha }\mathbf{e}_{\beta }-\
^{\alpha }\mathbf{e}_{\beta }\ ^{\alpha }\mathbf{e}_{\alpha }=\ ^{\alpha
}W_{\alpha \beta }^{\gamma }\ ^{\alpha }\mathbf{e}_{\gamma }.
\end{equation*}

\subsubsection{Fractional metrics and distinguished connections}

A (fractional) metric structure $\overset{\alpha }{\mathbf{g}}=\{\ ^{\alpha
}g_{\underline{\alpha }\underline{\beta }}\}$ is defined on a $\overset{%
\alpha }{\mathbf{V}}$ \ by a symmetric second rank tensor
\begin{equation}
\overset{\alpha }{\mathbf{g}}=\ ^{\alpha }g_{\underline{\gamma }\underline{%
\beta }}(u)(du^{\underline{\gamma }})^{\alpha }\otimes (du^{\underline{\beta
}})^{\alpha }.  \label{fmcf}
\end{equation}%
For N--adapted constructions, it is important to use the property that any
fractional metric $\overset{\alpha }{\mathbf{g}}$ can be represented
equivalently as a distinguished metric (d--metric), $\ \overset{\alpha }{%
\mathbf{g}}=\left[ \ ^{\alpha }g_{kj},\ ^{\alpha }g_{cb}\right] ,$ when
\begin{eqnarray}
\ \overset{\alpha }{\mathbf{g}} &=&\ ^{\alpha }g_{kj}(x,y)\ ^{\alpha
}e^{k}\otimes \ ^{\alpha }e^{j}+\ ^{\alpha }g_{cb}(x,y)\ ^{\alpha }\mathbf{e}%
^{c}\otimes \ ^{\alpha }\mathbf{e}^{b}  \label{m1} \\
&=&\eta _{k^{\prime }j^{\prime }}\ ^{\alpha }e^{k^{\prime }}\otimes \
^{\alpha }e^{j^{\prime }}+\eta _{c^{\prime }b^{\prime }}\ ^{\alpha }\mathbf{e%
}^{c^{\prime }}\otimes \ ^{\alpha }\mathbf{e}^{b^{\prime }},  \label{m1a}
\end{eqnarray}%
where matrices $\eta _{k^{\prime }j^{\prime }}=diag[\pm 1,\pm 1,...,\pm 1]$
and $\eta _{a^{\prime }b^{\prime }}=diag[\pm 1,\pm 1,...,\pm 1] $
(reflecting signature of ''prime'' spacetime $\mathbf{V}$) are obtained by
frame transforms
\begin{equation}
\eta _{k^{\prime }j^{\prime }}=e_{\ k^{\prime }}^{k}\ e_{\ j^{\prime }}^{j}\
_{\ }^{\alpha }g_{kj}\mbox{ and }\eta _{a^{\prime }b^{\prime }}=e_{\
a^{\prime }}^{a}\ e_{\ b^{\prime }}^{b}\ _{\ }^{\alpha }g_{ab}.
\label{mdecom}
\end{equation}%
For fractional computations, it is convenient to work with constants $\eta
_{k^{\prime }j^{\prime }}$ and $\eta _{a^{\prime }b^{\prime }}$ because the
Caputo derivatives of constants are zero. This allows us to keep the same
tensor rules as for the integer dimensions even the rules for taking local
derivatives became more sophisticate because of N--coefficients $\ ^{\alpha
}N_{i}^{a}(u)$ and additional vierbein transforms $e_{\ k^{\prime }}^{k}(u)$
and $e_{\ a^{\prime }}^{a}(u).$ Such coefficients mix fractional derivatives
$\overset{\alpha }{\underline{\partial }}_{a}$ computed as a local
integration (\ref{lfcd}). If we work with RL fractional derivatives, the
computation become very sophisticate with nonlinear mixing of integration,
partial derivatives etc.

A distinguished connection (d--connection) $\overset{\alpha }{\mathbf{D}}$
on $\overset{\alpha }{\mathbf{V}}$ is a linear connection preserving under
parallel transports the Whitney sum (\ref{whit}). Using the formalism of
fractional differential forms introduced in previous section, we can
elaborate a covariant fractional N--adapted calculus on nonholonomic
manifolds. \ To a fractional d--connection $\overset{\alpha }{\mathbf{D}}$ \
we can associate a N--adapted differential 1--form of type (\ref{fr1f})
\begin{equation}
\ ^{\alpha }\mathbf{\Gamma }_{\ \beta }^{\tau }=\ ^{\alpha }\mathbf{\Gamma }%
_{\ \beta \gamma }^{\tau }\ ^{\alpha }\mathbf{e}^{\gamma },  \label{fdcf}
\end{equation}%
when the coefficients are computed with respect to (\ref{ddif}) and (\ref%
{dder}) and para\-met\-rized $\ $\ the form $\ ^{\alpha }\mathbf{\Gamma }_{\
\tau \beta }^{\gamma }=\left( \ ^{\alpha }L_{jk}^{i},\ ^{\alpha
}L_{bk}^{a},\ ^{\alpha }C_{jc}^{i},\ ^{\alpha }C_{bc}^{a}\right) .$

On fractional forms $\overset{\alpha }{\mathbf{V}},$ we can act with the
absolute fractional differential $\ ^{\alpha }\mathbf{d}=\ _{\ _{1}x}\overset%
{\alpha }{d}_{x}+\ _{\ _{1}y}\overset{\alpha }{d}_{y}.$ In N--adapted
fractional form, the value $\ ^{\alpha }\mathbf{d:=}\ ^{\alpha }\mathbf{e}%
^{\beta }\ ^{\alpha }\mathbf{e}_{\beta }$ consists from exterior h- and
v--derivatives of type (\ref{feder}), \ when \
\begin{equation*}
\ _{\ _{1}x}\overset{\alpha }{d}_{x}:=(dx^{i})^{\alpha }\ \ _{\ _{1}x}%
\overset{\alpha }{\underline{\partial }}_{i}=\ ^{\alpha }e^{j}\ ^{\alpha }%
\mathbf{e}_{j}\mbox{ and }_{\ _{1}y}\overset{\alpha }{d}_{y}:=(dy^{a})^{%
\alpha }\ \ _{\ _{1}x}\overset{\alpha }{\underline{\partial }}_{a}=\
^{\alpha }\mathbf{e}^{b}\ ^{\alpha }e_{b}.
\end{equation*}

\subsubsection{Torsion and curvature of fractional d--connections}

The torsion and curvature of a fractional d--connection $\overset{\alpha }{%
\mathbf{D}}=\{\ ^{\alpha }\mathbf{\Gamma }_{\ \beta \gamma }^{\tau }\}$ are
computed, respectively, as fractional 2--forms,
\begin{eqnarray}
\ ^{\alpha }\mathcal{T}^{\tau }&\doteqdot& \overset{\alpha }{\mathbf{D}}\
^{\alpha }\mathbf{e}^{\tau }=\ ^{\alpha }\mathbf{d}\ ^{\alpha }\mathbf{e}%
^{\tau }+\ ^{\alpha }\mathbf{\Gamma }_{\ \beta }^{\tau }\wedge \ ^{\alpha }%
\mathbf{e}^{\beta } \mbox{ and }  \label{tors} \\
\ ^{\alpha }\mathcal{R}_{~\beta }^{\tau }&\doteqdot & \overset{\alpha }{%
\mathbf{D}}\mathbf{\ ^{\alpha }\Gamma }_{\ \beta }^{\tau }=\ ^{\alpha }%
\mathbf{d\ ^{\alpha }\Gamma }_{\ \beta }^{\tau }-\ ^{\alpha }\mathbf{\Gamma }%
_{\ \beta }^{\gamma}\wedge\ ^{\alpha }\mathbf{\Gamma }_{\ \gamma }^{\tau }=\
^{\alpha}\mathbf{R}_{\ \beta \gamma \delta}^{\tau}\ ^{\alpha}\mathbf{e}%
^{\gamma}\wedge\ ^{\alpha}\mathbf{e}.  \notag
\end{eqnarray}

The fractional Ricci tensor $\ ^{\alpha }\mathcal{R}ic=\{\ ^{\alpha }\mathbf{%
R}_{\alpha \beta }\doteqdot \ ^{\alpha }\mathbf{R}_{\ \alpha \beta
\tau}^{\tau }\}$ is
\begin{equation}
\ ^{\alpha }R_{ij}\doteqdot \ ^{\alpha }R_{\ ijk}^{k},\ \ \ ^{\alpha
}R_{ia}\doteqdot -\ ^{\alpha }R_{\ ika}^{k},\ \ ^{\alpha }R_{ai}\doteqdot \
^{\alpha }R_{\ aib}^{b},\ \ ^{\alpha }R_{ab}\doteqdot \ ^{\alpha }R_{\
abc}^{c}.  \label{dricci}
\end{equation}

The scalar curvature of a fractional d--connection $\overset{\alpha }{%
\mathbf{D}}$ is
\begin{equation*}
\ _{s}^{\alpha }\mathbf{R} \doteqdot \ ^{\alpha }\mathbf{g}^{\tau \beta }\
^{\alpha }\mathbf{R}_{\tau \beta }=\ ^{\alpha }R+\ ^{\alpha }S,\ ^{\alpha }R
=\ ^{\alpha }g^{ij}\ ^{\alpha }R_{ij},\ \ ^{\alpha }S=\ ^{\alpha }g^{ab}\
^{\alpha }R_{ab},
\end{equation*}
defined by a sum the h-- and v--components of (\ref{dricci}) and
contractions with the inverse coefficients to a d--metric (\ref{m1}).

We can introduce the Einstein tensor $\ ^{\alpha }\mathcal{E}ns=\{\ ^{\alpha
}\mathbf{G}_{\alpha \beta }\},$
\begin{equation}
\ ^{\alpha }\mathbf{G}_{\alpha \beta }:=\ ^{\alpha }\mathbf{R}_{\alpha \beta
}-\frac{1}{2}\ ^{\alpha }\mathbf{g}_{\alpha \beta }\ \ _{s}^{\alpha }\mathbf{%
R.}  \label{enstdt}
\end{equation}%
This allows us to elaborate various types of fractional models of gravity
(for different types of d--connections and fractional matter sources) \
generalizing the Einstein gravity theory and various modifications.

Finally, we note that for integer values of $\alpha$ the above formulas
transform into similar ones for nonholonomic manifolds \cite%
{ijgmmp,vrflg,vexsol}.

\subsubsection{Preferred fractional linear connections}

For applications in modern geometry and standard models of physics, there
are considered more special classes of d--connections:

\begin{itemize}
\item On a fractional nonholonomic $\overset{\alpha }{\mathbf{V}},$ there is
a unique canonical fractional d--connection $\ ^{\alpha }\widehat{\mathbf{D}}%
=\{\ ^{\alpha }\widehat{\mathbf{\Gamma }}_{\ \alpha \beta }^{\gamma }=\left(
\ ^{\alpha }\widehat{L}_{jk}^{i},\ ^{\alpha }\widehat{L}_{bk}^{a},\ ^{\alpha
}\widehat{C}_{jc}^{i},\ ^{\alpha }\widehat{C}_{bc}^{a}\right) \}$ which is
compatible with the metric structure, $\ ^{\alpha }\widehat{\mathbf{D}}\
\left( \ ^{\alpha }\mathbf{g}\right) =0,$ and satisfies the conditions $\
^{\alpha }\widehat{T}_{\ jk}^{i}=0$ and $\ ^{\alpha }\widehat{T}_{\
bc}^{a}=0.$\footnote{%
The N--adapted coefficients are explicitly determined by the (\ref{m1}),
\begin{eqnarray*}
\ ^{\alpha }\widehat{L}_{jk}^{i} &=&\frac{1}{2}\ ^{\alpha }g^{ir}\left(\
^{\alpha }\mathbf{e}_{k}\ ^{\alpha }g_{jr}+\ ^{\alpha }\mathbf{e}_{j}\
^{\alpha }g_{kr}-\ ^{\alpha }\mathbf{e}_{r}\ ^{\alpha }g_{jk}\right) ,
\label{candcon} \\
\ ^{\alpha }\widehat{L}_{bk}^{a} &=&\ ^{\alpha }e_{b}(\ ^{\alpha }N_{k}^{a})+%
\frac{1}{2}\ ^{\alpha }g^{ac}\left( \ ^{\alpha }\mathbf{e}_{k}\ ^{\alpha
}g_{bc}-\ ^{\alpha }g_{dc}\ \ ^{\alpha }e_{b}\ ^{\alpha }N_{k}^{d}-\
^{\alpha }g_{db}\ \ ^{\alpha }e_{c}\ ^{\alpha }N_{k}^{d}\right),  \notag \\
\ ^{\alpha }\widehat{C}_{jc}^{i} &=&\frac{1}{2}\ ^{\alpha }g^{ik}\ ^{\alpha
}e_{c}\ ^{\alpha }g_{jk}, \ \ ^{\alpha }\widehat{C}_{bc}^{a} =\frac{1}{2}\
^{\alpha }g^{ad}\left( \ ^{\alpha }e_{c}\ ^{\alpha }g_{bd}+\ ^{\alpha
}e_{c}\ ^{\alpha }g_{cd}-\ ^{\alpha }e_{d}\ ^{\alpha }g_{bc}\right) .  \notag
\end{eqnarray*}%
We can verify that introducing above formulas into (\ref{tors}) we obtain
that $\widehat{T}_{\ jk}^{i}=0$ and $\widehat{T}_{\ bc}^{a}=0,$ but $%
\widehat{T}_{\ ja}^{i},\widehat{T}_{\ ji}^{a}$ and $\widehat{T}_{\ bi}^{a}$
are not zero, and that the metricity conditions are satisfied in component
form.}

\item The Levi--Civita connection $\ ^{\alpha }\nabla =\{\ \ ^{\alpha
}\Gamma _{\ \alpha \beta }^{\gamma }\}$ can be defined in standard from but
by using the fractional Caputo left derivatives acting the coefficients of a
fractional metric (\ref{fmcf}).
\end{itemize}

As a consequence of nonholonomic structure, it is preferred to work on $%
\overset{\alpha }{\mathbf{V}}$ with $\ ^{\alpha }\widehat{\mathbf{D}}=\{\
^{\alpha }\widehat{\mathbf{\Gamma }}_{\ \tau \beta }^{\gamma }\}$ instead of
$\ ^{\alpha }\nabla $ (the last one is not adapted to the N--connection
splitting (\ref{whit})). The torsion $\ ^{\alpha }\widehat{\mathcal{T}}%
^{\tau }$ (\ref{tors}) \ of $\ ^{\alpha }\widehat{\mathbf{D}}$ is uniquely
induced nonholonomically by off--diagonal coefficients of the d--metric (\ref%
{m1}).

With respect to N--adapted fractional bases (\ref{dder}) and (\ref{ddif}),
the coefficients of the fractional Levi--Civita and canonical d--connection
satisfy the distorting relations\footnote{%
the N--adapted coefficients of distortion tensor $\ Z_{\ \alpha \beta
}^{\gamma }$ \ are computed%
\begin{eqnarray*}
\ \ \ ^{\alpha }Z_{jk}^{i} &=&0,\ \ \ ^{\alpha }Z_{jk}^{a}=-\ \ ^{\alpha
}C_{jb}^{i}\ \ ^{\alpha }g_{ik}\ \ ^{\alpha }g^{ab}-\frac{1}{2}\ \ ^{\alpha
}\Omega _{jk}^{a}, \\
\ \ \ ^{\alpha }Z_{bk}^{i} &=&\frac{1}{2}\ \ ^{\alpha }\Omega _{jk}^{c}\ \
^{\alpha }g_{cb}\ \ ^{\alpha }g^{ji}-\frac{1}{2}(\delta _{j}^{i}\delta
_{k}^{h}-\ \ ^{\alpha }g_{jk}\ \ ^{\alpha }g^{ih})\ \ ^{\alpha }C_{hb}^{j},
\\
\ \ \ ^{\alpha }Z_{bk}^{a} &=&\frac{1}{2}(\delta _{c}^{a}\delta _{d}^{b}+\ \
^{\alpha }g_{cd}\ \ ^{\alpha }g^{ab})\left[ \ \ ^{\alpha }L_{bk}^{c}-\ \
^{\alpha }e_{b}(\ \ ^{\alpha }N_{k}^{c})\right] , \\
\ \ \ ^{\alpha }Z_{kb}^{i} &=&\frac{1}{2}\ \ ^{\alpha }\Omega _{jk}^{a}\ \
^{\alpha }g_{cb}\ \ ^{\alpha }g^{ji}+\frac{1}{2}(\delta _{j}^{i}\delta
_{k}^{h}-\ \ ^{\alpha }g_{jk}\ \ ^{\alpha }g^{ih})\ \ ^{\alpha }C_{hb}^{j},
\\
\ ^{\alpha }Z_{jb}^{a} &=&-\frac{1}{2}(\delta _{c}^{a}\delta _{b}^{d}-\ \
^{\alpha }g_{cb}\ \ ^{\alpha }g^{ad})\left[\ ^{\alpha }L_{dj}^{c}-\ \
^{\alpha }e_{d}(\ \ ^{\alpha }N_{j}^{c})\right] , \ ^{\alpha }Z_{bc}^{a}=0,
\\
\ ^{\alpha }Z_{ab}^{i} &=&-\frac{\ \ ^{\alpha }g^{ij}}{2}\{\left[\ ^{\alpha
}L_{aj}^{c}-\ ^{\alpha }e_{a}(\ ^{\alpha }N_{j}^{c})\right]\ ^{\alpha
}g_{cb}+\left[\ ^{\alpha }L_{bj}^{c}-\ ^{\alpha }e_{b}(\ ^{\alpha }N_{j}^{c})%
\right]\ ^{\alpha }g_{ca}\}.
\end{eqnarray*}%
}
\begin{equation}
\ ^{\alpha }\Gamma _{\ \alpha \beta }^{\gamma }=\ ^{\alpha }\widehat{\mathbf{%
\Gamma }}_{\ \alpha \beta }^{\gamma }+\ \ ^{\alpha }Z_{\ \alpha \beta
}^{\gamma }.  \label{cdeft}
\end{equation}%
It is not possible to get relations of type (\ref{cdeft}) if the fractional
integro--differential structure would be not elaborated in N--adapted form
for the left Caputo derivative.

\subsection{Fractional Einstein equations for connections $\widehat{\mathbf{D%
}}$ and $\protect\nabla $}

In this section, we show that the fractional gravitational equations with
Caputo fractional derivatives can be integrated in general form similarly to
the results for integer dimensions \cite{vncg,ijgmmp,vsgg,vrflg,vexsol}.

\subsubsection{Nonholonomic variables in general relativity}

The Einstein equations on a spacetime manifold of integer dimension $\mathbf{%
V,}$ for an energy--momentum source of matter $T_{\alpha \beta },$ are
written in the form%
\begin{equation}
E_{\beta \gamma }=R_{\ \beta \delta }-\frac{1}{2}g_{\beta \delta
}R=\varkappa T_{\beta \delta },  \label{einsteq}
\end{equation}%
where $\varkappa =const$ and the Einstein tensor is computed for the
Levi--Civita connection $\nabla .$ It is not possible to integrate
 in any general form such nonlinear systems of partial
differential equations.

The Einstein equations (\ref{einsteq}) can be rewritten equivalently using
the canonical d--connection $\widehat{\mathbf{D}}=\{\widehat{\mathbf{\Gamma }%
}_{\ \alpha \beta }^{\gamma }\},$%
\begin{eqnarray}
\widehat{\mathbf{E}}_{\beta \delta }&=&\widehat{\mathbf{R}}_{\beta \delta }-%
\frac{1}{2}\mathbf{g}_{\beta \delta }\ ^{s}R=\mathbf{\Upsilon }_{\beta
\delta },  \label{cdeinst} \\
&&\widehat{L}_{aj}^{c}=e_{a}(N_{j}^{c}),\ \widehat{C}_{jb}^{i}=0,\ \Omega
_{\ ji}^{a}=0,  \label{lcconstr}
\end{eqnarray}%
where $\widehat{\mathbf{R}}_{\ \beta \delta }$ is the Ricci tensor for $%
\widehat{\mathbf{\Gamma }}_{\ \alpha \beta }^{\gamma },\ ^{s}R=\mathbf{g}%
^{\beta \delta }\widehat{\mathbf{R}}_{\ \beta \delta }$ and $\mathbf{%
\Upsilon }_{\beta \delta }$ is such a way constructed that $\mathbf{\Upsilon
}_{\beta \delta }\rightarrow \varkappa T_{\beta \delta }$ for $\widehat{%
\mathbf{D}}\rightarrow \nabla .$ In general, the Einstein tensor $\widehat{%
\mathbf{E}}_{\ \beta \delta }$ in (\ref{cdeinst}) is not equal to the
Einstein tensor $E_{\beta \gamma }$ in (\ref{einsteq}).

There are two possibilities to make equivalent two different systems of
equations for $\nabla $ and, respectively, for $\widehat{\mathbf{D}}.$ In
the first case, we can include the contributions of distortion tensor $Z_{\
\alpha \beta }^{\gamma }$ from (\ref{cdeft}) into the source $\mathbf{%
\Upsilon }_{\beta \delta }\sim \varkappa T_{\beta \delta }+\ ^{z}\mathbf{%
\Upsilon }_{\beta \delta }[Z_{\ \alpha \beta }^{\gamma }],$ in such a form
that the system (\ref{cdeinst}) is equivalent to (\ref{einsteq}) (both types
of such systems of equations are for the same metric structure $\mathbf{g}%
_{\beta \delta }$ but in terms of different connections). In the second
case, we consider that $\mathbf{\Upsilon }_{\beta \delta }=\varkappa
T_{\beta \delta }$ but in order to keep fundamental the Einstein equations
for $\nabla $ (even for some purposes we shall prefer to work with $\widehat{%
\mathbf{D}}$) we have to impose (at some final steps) the constraints (\ref%
{lcconstr}) when the tensors $\widehat{\mathbf{T}}_{\ \alpha \beta }^{\gamma
}$(\ref{tors}) and $Z_{\ \alpha \beta }^{\gamma }$(\ref{cdeft}) are zero.
For such constraints, we have $\ \widehat{\mathbf{\Gamma }}_{\ \alpha \beta
}^{\gamma }=\Gamma _{\ \alpha \beta }^{\gamma },$ with respect to N--adapted
frames (\ref{dder}) and (\ref{ddif}), even $\widehat{\mathbf{D}}\neq \nabla
. $

It is very surprising that for some general ansatz for metrics (see below
the integer version of (\ref{gsol})) there is a separation of equations (\ref%
{cdeinst}) for $\widehat{\mathbf{D}}$ which allows us to integrate such
systems in very general forms. If we work from the very beginning with $%
\nabla ,$ it is not possible to get from (\ref{einsteq}) a generally
solvable system of equations. Our idea was to encode the geometric and
physical data for $\nabla $ into $\widehat{\mathbf{D}}$ and work with
N--adapted constructions when, for instance, certain general solutions can
be found ''easily'' for various types of non--Riemannian and/or
Lagrange--Finsler theories. Imposing additional constraints (\ref{lcconstr}%
), we are able to extract Levi--Civita configurations with $\nabla ,$ for
instance, in Einstein gravity. This way, see original results and reviews in
a series of our works \cite{vncg,ijgmmp,vsgg,vrflg,vexsol}, we elaborated a
new method of constructing exact solutions with generic off--diagonal
metrics in gravity theories (the so--called anholonomic deformation/frame
method). Such a geometric technique seems also to be very efficient in Ricci
flow theories \cite{vnhrf1,vnhrf2,vnhrns,vnhrfs1,vnhrfs2,vrfnc,vfrrf} and
for elaborating different methods of quantization for gravity \cite%
{vdqlf,vbrane,avdqhc,vdqla,vqgr1,vqgr2}.

\subsubsection{Nonholonomic variables and fractional gravity}

Introducing the fractional canonical d--connection $\ ^{\alpha }\widehat{%
\mathbf{D}}$ into the Einstein d--tensor (\ref{enstdt}), following the same
principle of constructing the matter source $\ ^{\alpha }\mathbf{\Upsilon }%
_{\beta \delta }$ as in general relativity but for fractional
d--connections, we derive geometrically a fractional generalization of
N--adapted equations (\ref{cdeinst}),%
\begin{equation}
\ ^{\alpha }\widehat{\mathbf{E}}_{\ \beta \delta }=\ ^{\alpha }\mathbf{%
\Upsilon }_{\beta \delta }.  \label{fdeq}
\end{equation}%
Such a system can be restricted to fractional nonholonomic configurations
for $\ ^{\alpha }\nabla $ if we impose a fractional analog of constraints of
type (\ref{lcconstr})%
\begin{equation}
\ ^{\alpha }\widehat{L}_{aj}^{c}=\ ^{\alpha }e_{a}(\ ^{\alpha }N_{j}^{c}),\
\ ^{\alpha }\widehat{C}_{jb}^{i}=0,\ \ ^{\alpha }\Omega _{\ ji}^{a}=0.
\label{frconstr}
\end{equation}

There are not theoretical or experimental evidences that for integer
dimensions we must impose conditions of type (\ref{frconstr}). Nevertheless,
we shall consider them in section \ref{sbhfs} for deriving fractional black
hole solutions which will mimic maximally similar ones in general relativity.

\subsubsection{Separation of equations for fractional and integer dimensions}

One of the main purposes of this work is to prove that a very general ansatz
of type (\ref{m1}) defines exact solutions of the fractional Einstein
equations (\ref{fdeq}) metric. Let us consider a fractional metric
\begin{eqnarray}
\ _{\eta }^{\alpha }\mathbf{g} &\mathbf{=}&\mathbf{\ }^{\alpha }\eta
_{i}(x^{k},v)\ \ _{\circ }^{\alpha }g_{i}(x^{k},t)\mathbf{\ }^{\alpha }{%
dx^{i}\otimes \mathbf{\ }^{\alpha }dx^{i}}  \label{gsol} \\
&&+\mathbf{\ }^{\alpha }\eta _{a}(x^{k},v)\ \ _{\circ }^{\alpha
}h_{a}(x^{k},v)\mathbf{\ }^{\alpha }\mathbf{e}^{a}{\otimes }\mathbf{\ }%
^{\alpha }\mathbf{e}^{a},  \notag \\
\mathbf{\ }^{\alpha }\mathbf{e}^{3} &=&\mathbf{\ }^{\alpha }dv+\mathbf{\ }%
^{\alpha }\eta _{i}^{3}(x^{k},v)\ \ _{\circ }^{\alpha }w_{i}(x^{k},v)\mathbf{%
\ }^{\alpha }dx^{i},\   \notag \\
\mathbf{\ }^{\alpha }\mathbf{e}^{4} &=&\mathbf{\ }^{\alpha }dy^{4}+\mathbf{\
}^{\alpha }\eta _{i}^{4}(x^{k},v)\ \ _{\circ }^{\alpha }n_{i}(x^{k},v)%
\mathbf{\ }^{\alpha }dx^{i},  \notag
\end{eqnarray}%
when the coefficients will be defined below. For simplicity, we shall work
with the ''prime'' dimension splitting of type $2+2$ when coordinated are
labeled in the form $u^{\beta }=(x^{j},y^{3}=v,y^{4}),$ for $i,j,...=1,2.$
This ansatz has one Killing symmetry because the coefficients do not depend
explicitly on variable $y^{4}.$

In brief, we can write such the metric (\ref{gsol}) in the form
\begin{equation}
\ \mathbf{g}=\ ^{\alpha }g_{ij}\ ^{\alpha }{dx^{i}\otimes \ ^{\alpha }dx^{j}}%
+\ ^{\alpha }h_{ab}(\mathbf{\ }^{\alpha }dy^{a}+\mathbf{\ }^{\alpha
}N_{k}^{a}\ ^{\alpha }dx^{k}){\otimes}(\mathbf{\ }^{\alpha }dy^{b}+\
^{\alpha }N_{k}^{b}\mathbf{\ }^{\alpha }dx^{k}),  \label{dm}
\end{equation}%
where $\mathbf{\ }^{\alpha }g_{ij}=diag[\mathbf{\ }^{\alpha }g_{i}=\mathbf{\
}^{\alpha }\eta _{i}\ \ \ _{\circ }^{\alpha }g_{i}]$ and $\mathbf{\ }%
^{\alpha }h_{ab}=diag[\mathbf{\ }^{\alpha }h_{a}=\mathbf{\ }^{\alpha }\eta
_{a}\ \ \ _{\circ }^{\alpha }h_{a}]$ and $\mathbf{\ }^{\alpha }N_{k}^{3}=%
\mathbf{\ }^{\alpha }w_{i}=\mathbf{\ }^{\alpha }\eta _{i}^{3}\ \ \ _{\circ
}^{\alpha }w_{i}$ and $\mathbf{\ }^{\alpha }N_{k}^{4}=\mathbf{\ }^{\alpha
}n_{i}=\mathbf{\ }^{\alpha }\eta _{i}^{4}\ \ \ _{\circ }^{\alpha }n_{i}.$
The gravitational 'polarizations' $\mathbf{\ }^{\alpha }\eta _{\alpha }$ and
$\mathbf{\ }^{\alpha }\eta _{i}^{a}$ determine fractional nonholonomic
deformations of metrics, $\ \ _{\circ }^{\alpha }\mathbf{g}\mathbf{=}[\ \
_{\circ }^{\alpha }g_{i},\ \ _{\circ }^{\alpha }h_{a},\ \ _{\circ }^{\alpha
}N_{k}^{a}]\rightarrow \ \ _{\eta }^{\alpha }\mathbf{g}\mathbf{=}[\ \ _{\eta
}^{\alpha }g_{i},\ _{\eta }^{\alpha }h_{a},\ _{\eta }^{\alpha }N_{k}^{a}].$%
\footnote{%
Such transforms of geometric objects (with deformations of the frame,
metric, connections and other fundamental geometric structures) are more
general than those considered for the Cartan's moving frame method, when the
geometric objects are re--defined equivalently with respect to necessary
systems of reference.} \ The solutions of equations will be constructed for
a general source of type\footnote{%
such parametrizations of energy--momentum tensors are quite general ones for
\ various types of matter sources, which (in this work) are generalized for
fractional distributions; they can be defined by corresponding frame
transform}
\begin{equation}
\mathbf{\ }^{\alpha }\Upsilon _{\ \ \beta }^{\alpha }=diag[\mathbf{\ }%
^{\alpha }\Upsilon _{\gamma };\mathbf{\ }^{\alpha }\Upsilon _{1}=\mathbf{\ }%
^{\alpha }\Upsilon _{2}=\mathbf{\ }^{\alpha }\Upsilon _{2}(x^{k},v);\mathbf{%
\ }^{\alpha }\Upsilon _{3}=\mathbf{\ }^{\alpha }\Upsilon _{4}=\mathbf{\ }%
^{\alpha }\Upsilon _{4}(x^{k})]  \label{source}
\end{equation}

A straightforward computation\footnote{%
we omit such a cumbersome calculus which is similar to those presented in
Refs. \cite{ijgmmp,vsgg,vrflg,vexsol}; for the formulas considered in this work, we have to change usual partial N--adapted derivatives into fractional
ones considering that transform of type (\ref{mdecom}) might be performed in order to take into account the advantage that the action of Caputo derivative is zero for some constant coefficients} of the components of the
Ricci (\ref{dricci}) and Einstein (\ref{enstdt}) d--tensors corresponding to
ansatz (\ref{dm}) reduces the Einstein equations (\ref{fdeq}) to this system
of partial differential equations:
\begin{eqnarray}
\mathbf{\ }^{\alpha }\widehat{R}_{1}^{1} &=&\mathbf{\ }^{\alpha }\widehat{R}%
_{2}^{2}=-\frac{1}{2\mathbf{\ }^{\alpha }g_{1}\mathbf{\ }^{\alpha }g_{2}}%
\times [ \mathbf{\ }^{\alpha }g_{2}^{\bullet \bullet } -\frac{\mathbf{\ }%
^{\alpha }g_{1}^{\bullet }\mathbf{\ }^{\alpha }g_{2}^{\bullet }}{2\mathbf{\ }%
^{\alpha }g_{1}}  \label{eq1} \\
&&-\frac{\left( \mathbf{\ }^{\alpha }g_{2}^{\bullet }\right) ^{2}}{2\mathbf{%
\ }^{\alpha }g_{2}}+\mathbf{\ }^{\alpha }g_{1}^{\prime \prime }-\frac{%
\mathbf{\ }^{\alpha }g_{1}^{\prime }\mathbf{\ }^{\alpha }g_{2}^{\prime }}{2%
\mathbf{\ }^{\alpha }g_{2}}-\frac{\left( \mathbf{\ }^{\alpha
}g_{1}^{^{\prime }}\right) ^{2}}{2\mathbf{\ }^{\alpha }g_{1}}] =-\mathbf{\ }%
^{\alpha }\Upsilon _{4},  \notag \\
\mathbf{\ }^{\alpha }\widehat{R}_{3}^{3} &=&\mathbf{\ }^{\alpha }\widehat{R}%
_{4}^{4}=-\frac{1}{2\mathbf{\ }^{\alpha }h_{3}\mathbf{\ }^{\alpha }h_{4}}[%
\mathbf{\ }^{\alpha }h_{4}^{\ast \ast }  \label{eq2} \\
&&-\frac{\left( \mathbf{\ }^{\alpha }h_{4}^{\ast }\right) ^{2}}{2\mathbf{\ }%
^{\alpha }h_{4}}-\frac{\mathbf{\ }^{\alpha }h_{3}^{\ast }\mathbf{\ }^{\alpha
}h_{4}^{\ast }}{2\mathbf{\ }^{\alpha }h_{3}}] =-\mathbf{\ }^{\alpha
}\Upsilon _{2},  \notag
\end{eqnarray}

\begin{eqnarray}
\mathbf{\ }^{\alpha }\widehat{R}_{3k} &=&\frac{\mathbf{\ }^{\alpha }w_{k}}{2%
\mathbf{\ }^{\alpha }h_{4}}\left[ \mathbf{\ }^{\alpha }h_{4}^{\ast \ast }-%
\frac{\left( \mathbf{\ }^{\alpha }h_{4}^{\ast }\right) ^{2}}{2\mathbf{\ }%
^{\alpha }h_{4}}-\frac{\mathbf{\ }^{\alpha }h_{3}^{\ast }\mathbf{\ }^{\alpha
}h_{4}^{\ast }}{2\mathbf{\ }^{\alpha }h_{3}}\right]  \label{eq3} \\
&&+\frac{\mathbf{\ }^{\alpha }h_{4}^{\ast }}{4\mathbf{\ }^{\alpha }h_{4}}%
\left( \frac{_{\ _{1}x^{i}}\overset{\alpha }{\underline{\partial }}_{x^{i}}%
\mathbf{\ }^{\alpha }h_{3}}{\mathbf{\ }^{\alpha }h_{3}}+\frac{\partial _{k}%
\mathbf{\ }^{\alpha }h_{4}}{\mathbf{\ }^{\alpha }h_{4}}\right) -\frac{_{\
_{1}x^{k}}\overset{\alpha }{\underline{\partial }}_{x^{k}}\mathbf{\ }%
^{\alpha }h_{4}^{\ast }}{2\mathbf{\ }^{\alpha }h_{4}}=0,  \notag \\
\mathbf{\ }^{\alpha }\widehat{R}_{4k} &=&\frac{\mathbf{\ }^{\alpha }h_{4}}{2%
\mathbf{\ }^{\alpha }h_{3}}\mathbf{\ }^{\alpha }n_{k}^{\ast \ast }+\left(
\frac{\mathbf{\ }^{\alpha }h_{4}}{\mathbf{\ }^{\alpha }h_{3}}\mathbf{\ }%
^{\alpha }h_{3}^{\ast }-\frac{3}{2}\mathbf{\ }^{\alpha }h_{4}^{\ast }\right)
\frac{\mathbf{\ }^{\alpha }n_{k}^{\ast }}{2\mathbf{\ }^{\alpha }h_{3}}=0,
\label{eq4}
\end{eqnarray}%
In brief, we wrote the partial derivatives in the form
\begin{equation*}
\mathbf{\ }^{\alpha }a^{\bullet }=\overset{\alpha }{\underline{\partial }}%
_{1}a=_{\ _{1}x^{1}}\overset{\alpha }{\underline{\partial }}%
_{x^{1}}{}^{\alpha }a,\ \mathbf{\ }^{\alpha }a^{\prime }=\overset{\alpha }{%
\underline{\partial }}_{2}a=_{\ _{1}x^{2}}\overset{\alpha }{\underline{%
\partial }}_{x^{2}}{}^{\alpha }a,\ \ \mathbf{\ }^{\alpha }a^{\ast }=\overset{%
\alpha }{\underline{\partial }}_{v}a=_{\ _{1}v}\overset{\alpha }{\underline{%
\partial }}_{v}{}^{\alpha }a,
\end{equation*}%
$\ $\ see the left Caputo fractional derivatives (\ref{frlcb}).

Configurations with fractional Levi--Civita connection $\mathbf{\ }^{\alpha
}\nabla $ can be extracted by imposing additional constraints
\begin{eqnarray}
\mathbf{\ }^{\alpha }w_{i}^{\ast } &=&\mathbf{\ }^{\alpha }\mathbf{e}_{i}\ln
|\mathbf{\ }^{\alpha }h_{4}|,\mathbf{\ }^{\alpha }\mathbf{e}_{k}\mathbf{\ }%
^{\alpha }w_{i}=\mathbf{\ }^{\alpha }\mathbf{e}_{i}\mathbf{\ }^{\alpha
}w_{k},\ \   \notag \\
\mathbf{\ }^{\alpha }n_{i}^{\ast } &=&0,\ \overset{\alpha }{\underline{%
\partial }}_{i}\mathbf{\ }^{\alpha }n_{k}=\overset{\alpha }{\underline{%
\partial }}_{k}\mathbf{\ }^{\alpha }n_{i}  \label{frconstr1}
\end{eqnarray}%
satisfying the conditions (\ref{frconstr}).

Following the method considered in \cite{vexsol} for integer dimensions, we
can construct 'non--Killing' solutions depending on all coordinates when
\begin{eqnarray}
\mathbf{\ }^{\alpha }\mathbf{g} &\mathbf{=}&\mathbf{\ }^{\alpha }g_{i}(x^{k})%
\mathbf{\ }^{\alpha }{dx^{i}\otimes \mathbf{\ }^{\alpha }dx^{i}}+\mathbf{\ }%
^{\alpha }\omega ^{2}(x^{j},v,y^{4})\mathbf{\ }^{\alpha }h_{a}(x^{k},v)%
\mathbf{\ }^{\alpha }\mathbf{e}^{a}{\otimes }\mathbf{\ }^{\alpha }\mathbf{e}%
^{a},  \notag \\
\mathbf{\ }^{\alpha }\mathbf{e}^{3} &=&\mathbf{\ }^{\alpha }dy^{3}+\mathbf{\
}^{\alpha }w_{i}(x^{k},v)\mathbf{\ }^{\alpha }dx^{i},\mathbf{\ }^{\alpha }%
\mathbf{e}^{4}=\mathbf{\ }^{\alpha }dy^{4}+\mathbf{\ }^{\alpha
}n_{i}(x^{k},v)\mathbf{\ }^{\alpha }dx^{i},  \label{ansgensol}
\end{eqnarray}%
for any $\mathbf{\ }^{\alpha }\omega $ for which
\begin{equation*}
\mathbf{\ }^{\alpha }\mathbf{e}_{k}\mathbf{\ }^{\alpha }\omega =\overset{%
\alpha }{\underline{\partial }}_{k}\mathbf{\ }^{\alpha }\omega +\mathbf{\ }%
^{\alpha }w_{k}\mathbf{\ }^{\alpha }\omega ^{\ast }+\mathbf{\ }^{\alpha
}n_{k}\overset{\alpha }{\underline{\partial }}_{y^{4}}\mathbf{\ }^{\alpha
}\omega =0,
\end{equation*}%
when (\ref{ansgensol}) with $\mathbf{\ }^{\alpha }\omega ^{2}=1$ is of type (%
\ref{dm}). The length of this paper does not allow us to study such general
fractional solutions.

\section{General Solutions for Fractional Einstein Equations}

\label{sesfee}

There is a very important separation property in the system (\ref{eq1})--(\ref{eq4})
which allows us to find exact very general solutions for such equations both
for non--integer and integer dimensions. For instance, if the coefficient $%
\mathbf{\ }^{\alpha }g_{1}(x^{k})$ is known, we can find from (\ref{eq1})
the value of $\mathbf{\ }^{\alpha }g_{2}(x^{k})$ (we may consider an inverse
situation when $\mathbf{\ }^{\alpha }g_{1}(x^{k})$ is to be computed for a
known value of $\mathbf{\ }^{\alpha }g_{2}(x^{k})$). Similarly, we can
define from (\ref{eq2}) the value of $\ ^{\alpha }h_{3}(x^{k},v)$ from $\
^{\alpha }h_{4}(x^{k},v),$ or inversely. Having defined $\ ^{\alpha }h_{a},$ at the third step,
we  compute the N--connection coefficients: $\mathbf{\ }^{\alpha
}w_{i}(x^{k},v)$ are certain solutions of algebraic equations (\ref{eq3}).
 Then, we  compute $\mathbf{\ }^{\alpha }n_{i}(x^{k},v)$ by integrating two
times on $v$ in (\ref{eq4}).  Such a property of exact integration of the field equations can not obtained for
fractional gravity models with the RL derivative but exists for those based on the Caputo one, correspondingly adapted to some classes of nonholonomic distributions.

The explicit form of solutions of the fractional Einstein equations depend on the values of the coefficients in the ansatz for metric and source. Let us show how such solutions are constructed for ansatz of type (\ref{dm}).

\subsection{Solutions with $\mathbf{\ }^{\protect\alpha }h_{3,4}^{\ast }\neq
0$ and $\mathbf{\ }^{\protect\alpha }\Upsilon _{2,4}\neq 0$}

Such metrics are defined by ansatz
\begin{eqnarray}
\ \mathbf{\ }^{\alpha }\mathbf{g} &\mathbf{=}&e^{\mathbf{\ }^{\alpha }\psi
(x^{k})}\mathbf{\ }^{\alpha }{dx^{i}\otimes \mathbf{\ }^{\alpha }dx^{i}}%
+h_{3}(x^{k},v)\mathbf{\ }^{\alpha }\mathbf{e}^{3}{\otimes }\mathbf{\ }%
^{\alpha }\mathbf{e}^{3}+h_{4}(x^{k},v)\mathbf{\ }^{\alpha }\mathbf{e}^{4}{%
\otimes }\mathbf{\ }^{\alpha }\mathbf{e}^{4},  \notag \\
\mathbf{\ }^{\alpha }\mathbf{e}^{3} &=&\mathbf{\ }^{\alpha }dv+\mathbf{\ }%
^{\alpha }w_{i}(x^{k},v)\mathbf{\ }^{\alpha }dx^{i},\mathbf{\ }^{\alpha }%
\mathbf{e}^{4}=\mathbf{\ }^{\alpha }dy^{4}+\mathbf{\ }^{\alpha
}n_{i}(x^{k},v)\mathbf{\ }^{\alpha }dx^{i}  \label{genans}
\end{eqnarray}%
with the coefficients being solutions of the system
\begin{eqnarray}
\mathbf{\ }^{\alpha }\ddot{\psi}+\mathbf{\ }^{\alpha }\psi ^{\prime \prime }
&=&2\mathbf{\ }^{\alpha }\Upsilon _{4}(x^{k}),  \label{4ep1a} \\
\mathbf{\ }^{\alpha }h_{4}^{\ast } &=&2\mathbf{\ }^{\alpha }h_{3}\mathbf{\ }%
^{\alpha }h_{4}\mathbf{\ }^{\alpha }\Upsilon _{2}(x^{i},v)/\mathbf{\ }%
^{\alpha }\phi ^{\ast },  \label{4ep2a}
\end{eqnarray}%
\begin{eqnarray}
\mathbf{\ }^{\alpha }\beta \mathbf{\ }^{\alpha }w_{i}+\mathbf{\ }^{\alpha
}\alpha _{i} &=&0,  \label{4ep3a} \\
\mathbf{\ }^{\alpha }n_{i}^{\ast \ast }+\mathbf{\ }^{\alpha }\gamma \mathbf{%
\ }^{\alpha }n_{i}^{\ast } &=&0,  \label{4ep4a}
\end{eqnarray}%
\begin{eqnarray}
\mbox{where} \mathbf{\ }^{\alpha }\phi &=&\ln |\frac{\mathbf{\ }^{\alpha
}h_{4}^{\ast }}{\sqrt{|\mathbf{\ }^{\alpha }h_{3}\mathbf{\ }^{\alpha }h_{4}|}%
}|,\ \mathbf{\ }^{\alpha }\gamma =\left( \ln |\mathbf{\ }^{\alpha
}h_{4}|^{3/2}/|\mathbf{\ }^{\alpha }h_{3}|\right) ^{\ast },  \label{auxphi}
\\
\mathbf{\ }^{\alpha }\alpha _{i} &=&\mathbf{\ }^{\alpha }h_{4}^{\ast }%
\overset{\alpha }{\underline{\partial }}_{k}\ ^{\alpha }\phi ,\ \mathbf{\ }%
^{\alpha }\beta =\mathbf{\ }^{\alpha }h_{4}^{\ast }\ \mathbf{\ }^{\alpha
}\phi ^{\ast }\ .  \notag
\end{eqnarray}%
For $\mathbf{\ }^{\alpha }h_{4}^{\ast }\neq 0;\mathbf{\ }^{\alpha }\Upsilon
_{2}\neq 0,$ we have $\mathbf{\ }^{\alpha }\phi ^{\ast }\neq 0.$ \ The
exponential function $e^{\mathbf{\ }^{\alpha }\psi (x^{k})}$ in (\ref{genans}%
) is the fractional analog of the ''integer'' exponential functions and
called the Mittag--Leffer function $E_{\alpha }[(x-\ ^{1}x)^{\alpha }].$ For
$^{\alpha }\psi (x)=E_{\alpha }[(x-\ ^{1}x)^{\alpha }],$ we have $\overset{%
\alpha }{\underline{\partial }}_{i}E_{\alpha }=E_{\alpha },$ see (for
instance) \cite{taras08}. For simplicity, hereafter we shall write usual
symbols for functions as in the case of integer calculus, but providing a
label $\alpha $ considering such fractional construction as certain Taylor
series in \cite{lrz}.

It is possible to consider any nonconstant $\mathbf{\ }^{\alpha }\phi =%
\mathbf{\ }^{\alpha }\phi (x^{i},v)$ as a generating function, we can
construct exact solutions of (\ref{4ep1a})--(\ref{4ep4a}). We have to solve
respectively the two dimensional fractional Laplace equation, for $\ \mathbf{%
\ }^{\alpha }g_{1}=\ \mathbf{\ }^{\alpha }g_{2}=e^{\ \mathbf{\ }^{\alpha
}\psi (x^{k})}.$ Then we integrate on $v,$ in order to determine $\mathbf{\ }%
^{\alpha }h_{3},$ $\mathbf{\ }^{\alpha }h_{4}$ and $\mathbf{\ }^{\alpha
}n_{i},$ and solving algebraic equations, for $\mathbf{\ }^{\alpha }w_{i}.$
We obtain (computing consequently for a chosen $\mathbf{\ }^{\alpha }\phi
(x^{k},v)$)
\begin{eqnarray}
\mathbf{\ }^{\alpha }g_{1} &=&\mathbf{\ }^{\alpha }g_{2}=e^{\mathbf{\ }%
^{\alpha }\psi (x^{k})},\mathbf{\ }^{\alpha }h_{3}=\pm \ \frac{|\mathbf{\ }%
^{\alpha }\phi ^{\ast }(x^{i},v)|}{\mathbf{\ }^{\alpha }\Upsilon _{2}},\
\label{gsol1} \\
\mathbf{\ }^{\alpha }h_{4} &=&\ \mathbf{\ }_{0}^{\alpha }h_{4}(x^{k})\pm \
2_{\ _{1}v}\overset{\alpha }{I}_{v}\frac{(\exp [2\ \mathbf{\ }^{\alpha }\phi
(x^{k},v)])^{\ast }}{\mathbf{\ }^{\alpha }\Upsilon _{2}},\   \notag \\
\mathbf{\ }^{\alpha }w_{i} &=&-\overset{\alpha }{\underline{\partial }}_{i}%
\mathbf{\ }^{\alpha }\phi /\mathbf{\ }^{\alpha }\phi ^{\ast },  \notag \\
\mathbf{\ }^{\alpha }n_{i} &=&\ _{1}^{\alpha }n_{k}\left( x^{i}\right) +\
_{2}^{\alpha }n_{k}\left( x^{i}\right) _{\ _{1}v}\overset{\alpha }{I}_{v}[%
\mathbf{\ }^{\alpha }h_{3}/(\sqrt{|\mathbf{\ }^{\alpha }h_{4}|})^{3}],
\notag
\end{eqnarray}%
where $\ \mathbf{\ }_{0}^{\alpha }h_{4}(x^{k}),\ \mathbf{\ }_{1}^{\alpha
}n_{k}\left( x^{i}\right) $ and $\ \mathbf{\ }_{2}^{\alpha }n_{k}\left(
x^{i}\right) $ are integration functions, and $_{\ _{1}v}\overset{\alpha }{I}%
_{v}$ is the fractional integral on variables $v.$

Here we note that the solutions (\ref{gsol1}), and, in general, almost all
solutions in fractional calculus with left Caputo derivatives, can be considered as some series
expansions and relevant fractional differential equations as is sketched in
ref. \cite{lrz}. In such an approach, for various types of fractional
functions, there are defined a kind of fractional Taylor series of infinitely
fractionally--differentiable functions.

To construct exact solutions for the Levi--Civita connection $\mathbf{\ }%
^{\alpha }\nabla ,$ we have to constrain the coefficients (\ref{gsol1}) to
satisfy the conditions (\ref{frconstr1}). For instance, we can fix a
nonholonomic distribution when $\ \mathbf{\ }_{2}^{\alpha }n_{k}\left(
x^{i}\right) $ $=0$ and $\ _{1}^{\alpha }n_{k}\left( x^{i}\right) $ are any
functions satisfying the conditions $\ \overset{\alpha }{\underline{\partial
}}_{i}\mathbf{\ }_{1}^{\alpha }n_{k}\left( xj\right) =\overset{\alpha }{%
\underline{\partial }}_{k}\mathbf{\ }_{1}^{\alpha }n_{i}\left( x^{j}\right)
. $ The constraints on $\mathbf{\ }^{\alpha }\phi (x^{k},v)$ are related to
the N--connection coefficients $\mathbf{\ }^{\alpha }w_{i}=-\overset{\alpha }%
{\underline{\partial }}_{i}\mathbf{\ }^{\alpha }\phi /\ \mathbf{\ }^{\alpha
}\phi ^{\ast }$ following relations
\begin{eqnarray}
\left( \mathbf{\ }^{\alpha }w_{i}[\mathbf{\ }^{\alpha }\phi ]\right) ^{\ast
}+\mathbf{\ }^{\alpha }w_{i}[\mathbf{\ }^{\alpha }\phi ]\left( \mathbf{\ }%
^{\alpha }h_{4}[\mathbf{\ }^{\alpha }\phi ]\right) ^{\ast }+\overset{\alpha }%
{\underline{\partial }}_{i}\mathbf{\ }^{\alpha }h_{4}[\mathbf{\ }^{\alpha
}\phi ]=0, &&  \notag \\
\overset{\alpha }{\underline{\partial }}_{i}\mathbf{\ }^{\alpha }w_{k}[%
\mathbf{\ }^{\alpha }\phi ]=\overset{\alpha }{\underline{\partial }}_{k}\
\mathbf{\ }^{\alpha }w_{i}[\mathbf{\ }^{\alpha }\phi ], &&  \label{auxc1}
\end{eqnarray}%
where, for instance, we denoted by $\mathbf{\ }^{\alpha }h_{4}[\mathbf{\ }%
^{\alpha }\phi ]$ the functional dependence on $\mathbf{\ }^{\alpha }\phi .$
Such conditions are always satisfied for $\mathbf{\ }^{\alpha }\phi =\mathbf{%
\ }^{\alpha }\phi (v)$ or if $\mathbf{\ }^{\alpha }\phi =const$ \ when $%
\mathbf{\ }^{\alpha }w_{i}(x^{k},v)$ can be any functions as follows from (%
\ref{4ep3a}) with zero $\mathbf{\ }^{\alpha }\beta $ and $\mathbf{\ }%
^{\alpha }\alpha _{i},$ see (\ref{auxphi})).

\subsection{Three other important classes of solutions}

If any of the conditions $\mathbf{\ }^{\alpha }h_{3,4}^{\ast }\neq 0$ is not
satisfied, we can construct another types of solutions for certain special
parametrization of coefficients for ansatz (\ref{gsol1}) subjected to the
condition to be a solution of equations (\ref{4ep1a})--(\ref{4ep4a}).

\subsubsection{Solutions with $\mathbf{\ }^{\protect\alpha }h_{4}^{\ast }=0$}

The equation (\ref{eq2}) can be solved for such a case, $\mathbf{\ }^{\alpha
}h_{4}^{\ast }=0,$ only if $\mathbf{\ }^{\alpha }\Upsilon _{2}=0.$ Any set
of functions $\mathbf{\ }^{\alpha }w_{i}(x^{k},v)$ can define a solution of (%
\ref{eq3}), and its equivalent (\ref{4ep3a}), because the coefficients $%
\mathbf{\ }^{\alpha }\beta $ and $\mathbf{\ }^{\alpha }\alpha _{i},$ see (%
\ref{auxphi}), are zero. The coefficients $\mathbf{\ }^{\alpha }n_{i}$ are
determined from (\ref{4ep4a}) with $\mathbf{\ }^{\alpha }h_{4}^{\ast }=0$
and any given $\mathbf{\ }^{\alpha }h_{3}$ which results in $\ \mathbf{\ }%
^{\alpha }n_{k}=\ _{1}^{\alpha }n_{k}\left( x^{i}\right) +\ \mathbf{\ }%
_{2}^{\alpha }n_{k}\left( x^{i}\right) \ _{\ _{1}v}\overset{\alpha }{I}_{v}%
\mathbf{\ }^{\alpha }h_{3}.$ It is possible to take $\mathbf{\ }^{\alpha }$ $%
g_{1}=\mathbf{\ }^{\alpha }g_{2}=e^{\mathbf{\ }^{\alpha }\psi (x^{k})},$
with $\mathbf{\ }^{\alpha }\psi (x^{k})$ determined by (\ref{4ep1a}) for a
given $\mathbf{\ }^{\alpha }\Upsilon _{4}(x^{k}).$

This class of solutions is given by ansatz
\begin{eqnarray}
\ \mathbf{\ }^{\alpha }\mathbf{g} &\mathbf{=}&e^{\mathbf{\ }^{\alpha }\psi
(x^{k})\mathbf{\ }^{\alpha }}{dx^{i}\otimes \mathbf{\ }^{\alpha }dx^{i}}
\label{genans1} \\
&&+\mathbf{\ }^{\alpha }h_{3}(x^{k},v)\mathbf{\ }^{\alpha }\mathbf{e}^{3}{%
\otimes }\mathbf{\ }^{\alpha }\mathbf{e}^{3}+\ \mathbf{\ }_{0}^{\alpha
}h_{4}(x^{k})\mathbf{\ }^{\alpha }\mathbf{e}^{4}{\otimes }\mathbf{\ }%
^{\alpha }\mathbf{e}^{4},  \notag \\
{\mathbf{\ }^{\alpha }}\mathbf{e}^{3} &=&{\mathbf{\ }^{\alpha }}dv+{\mathbf{%
\ }^{\alpha }}w_{i}(x^{k},v){\mathbf{\ }^{\alpha }}dx^{i},  \notag \\
\ {\mathbf{\ }^{\alpha }}\mathbf{e}^{4} &=&{\mathbf{\ }^{\alpha }}dy^{4}+%
\left[ \ {\mathbf{\ }_{1}^{\alpha }}n_{k}\left( x^{i}\right) +\ {\mathbf{\ }%
_{2}^{\alpha }}n_{k}\left( x^{i}\right) \ _{\ _{1}v}\overset{\alpha }{I}%
_{v}h_{3}\right] {\mathbf{\ }^{\alpha }}dx^{i},  \notag
\end{eqnarray}%
for arbitrary generating fractional functions ${\mathbf{\ }^{\alpha }}%
h_{3}(x^{k},v),{\mathbf{\ }^{\alpha }}w_{i}(x^{k},v),\ {\mathbf{\ }%
_{0}^{\alpha }}h_{4}(x^{k})$ and integration fractional functions $\ {%
\mathbf{\ }_{1}^{\alpha }}n_{k}\left( x^{i}\right) $ and $\ {\mathbf{\ }%
_{2}^{\alpha }}n_{k}\left( x^{i}\right) .$

A subclass of solutions for the Levi--Civita connection can be selected from
(\ref{genans1}) by imposing the conditions (\ref{frconstr1})
\begin{eqnarray*}
\ {\mathbf{\ }_{2}^{\alpha }}n_{k}\left( x^{i}\right) =0\ &\mbox{ and }&%
\overset{\alpha }{\underline{\partial }}_{i}\mathbf{\ }{_{1}^{\alpha }}n_{k}=%
\overset{\alpha }{\underline{\partial }}_{k}\mathbf{\ }{_{1}^{\alpha }}n_{i},
\\
{\mathbf{\ }^{\alpha }}w_{i}^{\ast }+\overset{\alpha }{\underline{\partial }}%
_{i}\ ^{0}h_{4}=0 &\mbox{ and }&\overset{\alpha }{\underline{\partial }}_{i}{%
\mathbf{\ }^{\alpha }}w_{k}=\overset{\alpha }{\underline{\partial }}_{k}{%
\mathbf{\ }^{\alpha }}w_{i},
\end{eqnarray*}%
for any such ${\mathbf{\ }^{\alpha }}w_{i}(x^{k},v)$ and $\ {\mathbf{\ }%
_{0}^{\alpha }}h_{4}(x^{k}).$

\subsubsection{Solutions with ${\mathbf{\ }^{\protect\alpha }}h_{3}^{\ast
}=0 $ and ${\mathbf{\ }^{\protect\alpha }}h_{4}^{\ast }\neq 0$}

The ansatz for metric is of type
\begin{eqnarray}
\ \mathbf{\ }^{\alpha }\mathbf{g} &\mathbf{=}&e^{\mathbf{\ }^{\alpha }\psi
(x^{k})\mathbf{\ }^{\alpha }}{dx^{i}\otimes \mathbf{\ }^{\alpha }dx^{i}}
\notag \\
&&-\ {\mathbf{\ }_{0}^{\alpha }}h_{3}(x^{k})\mathbf{\ }^{\alpha }\mathbf{e}%
^{3}{\otimes }\mathbf{\ }^{\alpha }\mathbf{e}^{3}+\mathbf{\ }^{\alpha
}h_{4}(x^{k},v)\mathbf{\ }^{\alpha }\mathbf{e}^{4}{\otimes }\mathbf{\ }%
^{\alpha }\mathbf{e}^{4},  \notag \\
\mathbf{\ }^{\alpha }\mathbf{e}^{3} &=&\mathbf{\ }^{\alpha }dv+\mathbf{\ }%
^{\alpha }w_{i}(x^{k},v)\mathbf{\ }^{\alpha }dx^{i},  \label{genans2} \\
\mathbf{\ }^{\alpha }\mathbf{e}^{4} &=&\mathbf{\ }^{\alpha }dy^{4}+\mathbf{\
}^{\alpha }n_{i}(x^{k},v)\mathbf{\ }^{\alpha }dx^{i},  \notag
\end{eqnarray}%
where $\mathbf{\ }^{\alpha }g_{1}=\mathbf{\ }^{\alpha }g_{2}=e^{\mathbf{\ }%
^{\alpha }\psi (x^{k})},$ for $\mathbf{\ }^{\alpha }\psi (x^{k})$ being a
solution of (\ref{4ep1a}) with a given $\mathbf{\ }^{\alpha }\Upsilon
_{4}(x^{k}).$ A function $\mathbf{\ }^{\alpha }h_{4}(x^{k},v)$ solves the
equation (\ref{4ep2a}) for $\mathbf{\ }^{\alpha }h_{3}^{\ast }=0,$ which can
be represented
\begin{equation*}
\mathbf{\ }^{\alpha }h_{4}^{\ast \ast }-\frac{\left( \mathbf{\ }^{\alpha
}h_{4}^{\ast }\right) ^{2}}{2\mathbf{\ }^{\alpha }h_{4}}-2\ \mathbf{\ }%
_{0}^{\alpha }h_{3}\mathbf{\ }^{\alpha }h_{4}\mathbf{\ }^{\alpha }\Upsilon
_{2}(x^{k},v)=0.
\end{equation*}%
The solutions for the N--connection coefficients are
\begin{eqnarray*}
\mathbf{\ }^{\alpha }w_{i} &=&-\overset{\alpha }{\underline{\partial }}_{i}%
\mathbf{\ }^{\alpha }\widetilde{\phi }/\mathbf{\ }^{\alpha }\widetilde{\phi }%
^{\ast },~ \\
\mathbf{\ }^{\alpha }n_{i} &=&\ \mathbf{\ }_{1}^{\alpha }n_{k}\left(
x^{i}\right) +\ \mathbf{\ }_{2}^{\alpha }n_{k}\left( x^{i}\right) _{\ _{1}v}%
\overset{\alpha }{I}_{v}[1/(\sqrt{|\mathbf{\ }^{\alpha }h_{4}|})^{3}],
\end{eqnarray*}%
when $\mathbf{\ }^{\alpha }\widetilde{\phi }=\ln |\mathbf{\ }^{\alpha
}h_{4}^{\ast }/\sqrt{|\ \mathbf{\ }_{0}^{\alpha }h_{3}\mathbf{\ }^{\alpha
}h_{4}|}|.$

The Levi--Civita conditions (\ref{frconstr1}) for (\ref{genans2}) are
\begin{equation*}
\ \ \mathbf{\ }_{1}^{\alpha }n_{k}\left( x^{i}\right) =0\ \mbox{ and }%
\overset{\alpha }{\underline{\partial }}_{i}\mathbf{\ }_{1}^{\alpha }n_{k}=%
\overset{\alpha }{\underline{\partial }}_{k}\mathbf{\ }_{1}^{\alpha }n_{i},
\end{equation*}%
\begin{eqnarray*}
\left( \mathbf{\ }^{\alpha }w_{i}[\mathbf{\ }^{\alpha }\widetilde{\phi }%
]\right) ^{\ast }+\mathbf{\ }^{\alpha }w_{i}[\mathbf{\ }^{\alpha }\widetilde{%
\phi }]\left( \mathbf{\ }^{\alpha }h_{4}[\mathbf{\ }^{\alpha }\widetilde{%
\phi }]\right) ^{\ast }+\overset{\alpha }{\underline{\partial }}_{i}\mathbf{%
\ }^{\alpha }h_{4}[\mathbf{\ }^{\alpha }\widetilde{\phi }]=0, && \\
\overset{\alpha }{\underline{\partial }}_{i}\mathbf{\ }^{\alpha }w_{k}[%
\mathbf{\ }^{\alpha }\widetilde{\phi }]=\overset{\alpha }{\underline{%
\partial }}_{k}\mathbf{\ }^{\alpha }w_{i}[\mathbf{\ }^{\alpha }\widetilde{%
\phi }]. &&
\end{eqnarray*}%
For small fractional deformations, it is not obligatory to impose such
conditions. We can consider integer Levi--Civita configurations and then to
transform them nonholonomically into certain d--connection ones.

\subsubsection{Solutions with $\mathbf{\ }^{\protect\alpha }\protect\phi %
=const$}

Fixing in (\ref{auxphi}) $\mathbf{\ }^{\alpha }\phi =\mathbf{\ }^{\alpha
}\phi _{0}=const$ and considering $\mathbf{\ }^{\alpha }h_{3}^{\ast }\neq 0$
and $\mathbf{\ }^{\alpha }h_{4}^{\ast }\neq 0,$ we get that the general
solutions of (\ref{4ep1a})--(\ref{4ep4a}) are
\begin{eqnarray}
\ ^{\alpha }\mathbf{g} &=&e^{\mathbf{\ }^{\alpha }\psi (x^{k})\mathbf{\ }%
^{\alpha }}{dx^{i}\otimes \mathbf{\ }^{\alpha }dx^{i}}-  \notag \\
&&\ _{0}^{\alpha }h^{2}[\mathbf{\ }^{\alpha }f^{\ast }( x^{i},v)] ^{2}|%
\mathbf{\ }^{\alpha }\varsigma _{\Upsilon }\left( x^{i},v\right) |\mathbf{\ }%
^{\alpha }\mathbf{e}^{3}{\otimes }\mathbf{\ }^{\alpha }\mathbf{e}^{3} +%
\mathbf{\ }^{\alpha }f^{2}\left( x^{i},v\right) \mathbf{\ }^{\alpha }\mathbf{%
e}^{4}{\otimes }\mathbf{\ }^{\alpha }\mathbf{e}^{4},  \notag \\
\ \mathbf{\ }^{\alpha }\mathbf{e}^{3} &=&\mathbf{\ }^{\alpha }dv+ \ ^{\alpha
}w_{i}(x^{k},v)\mathbf{\ }^{\alpha }dx^{i},\ ^{\alpha }\mathbf{e}^{4}= \
^{\alpha }dy^{4}+ \ ^{\alpha }n_{k}\left( x^{i},v\right) \mathbf{\ }^{\alpha
}dx^{i},  \label{genans3}
\end{eqnarray}%
where $\ \mathbf{\ }_{0}^{\alpha }h=const$ and $\mathbf{\ }^{\alpha }g_{1}=%
\mathbf{\ }^{\alpha }g_{2}=e^{\mathbf{\ }^{\alpha }\psi (x^{k})},$ with $%
\mathbf{\ }^{\alpha }\psi (x^{k})$ being a solution of (\ref{4ep1a}) for any
given $\mathbf{\ }^{\alpha }\Upsilon _{4}(x^{k}).$

Using the fractional function
\begin{equation*}
\mathbf{\ }^{\alpha }\varsigma _{\Upsilon }\left( x^{i},v\right) =\mathbf{\ }%
^{\alpha }\varsigma _{4[0]}\left( x^{i}\right) -\frac{\ _{0}^{\alpha }h^{2}}{%
16}_{\ _{1}v}\overset{\alpha }{I}_{v}\Upsilon _{2}(x^{k},v)[\mathbf{\ }%
^{\alpha }f^{2}\left( x^{i},v\right) ]^{2}\ ,
\end{equation*}
we write the fractional solutions for N--connection coefficients $\ ^{\alpha
}N_{i}^{3}=\mathbf{\ }^{\alpha }w_{i}$ and $\ ^{\alpha }N_{i}^{4}=\mathbf{\ }%
^{\alpha }n_{i}$ in the form
\begin{eqnarray}
\mathbf{\ }^{\alpha }w_{i}&=&-\overset{\alpha }{\underline{\partial }}_{i}%
\mathbf{\ }^{\alpha }\varsigma _{\Upsilon }\left( x^{k},v\right) /\mathbf{\ }%
^{\alpha }\varsigma _{\Upsilon }^{\ast }\left( x^{k},v\right)
\label{gensol1w} \\
\mbox{and } \mathbf{\ }^{\alpha }n_{k}&=&\ _{1}^{\alpha }n_{k}\left(
x^{i}\right) +\ _{2}^{\alpha }n_{k}\left( x^{i}\right) _{\ _{1}v}\overset{%
\alpha }{I}_{v}\frac{\left[ \mathbf{\ }^{\alpha }f^{\ast }\left(
x^{i},v\right) \right] ^{2}}{\left[ \mathbf{\ }^{\alpha }f\left(
x^{i},v\right) \right] ^{2}}\mathbf{\ }^{\alpha }\varsigma _{\Upsilon
}\left( x^{i},v\right).  \label{gensol1n}
\end{eqnarray}

If $\mathbf{\ }^{\alpha }\varsigma _{\Upsilon }\left( x^{i},v\right) =\pm 1$
for $\mathbf{\ }^{\alpha }\Upsilon _{2}\rightarrow 0,$ we take $\mathbf{\ }%
^{\alpha }\varsigma _{4[0]}\left( x^{i}\right) =\pm 1.$ For such conditions,
the functions $\mathbf{\ }^{\alpha }h_{3}=-\ \mathbf{\ }_{0}^{\alpha }h^{2}\ %
\left[ \mathbf{\ }^{\alpha }f^{\ast }\left( x^{i},v\right) \right] ^{2}$ and
$\mathbf{\ }^{\alpha }h_{4}=\mathbf{\ }^{\alpha }f^{2}\left( x^{i},v\right) $
satisfy the equation (\ref{4ep2a}), when $\sqrt{|\mathbf{\ }^{\alpha }h_{3}|}%
=\ ^{0}h(\sqrt{|\mathbf{\ }^{\alpha }h_{4}|})^{\ast }$ is compatible with
the condition $\mathbf{\ }^{\alpha }\phi =\mathbf{\ }^{\alpha }\phi _{0}.$

The subclass of solutions for the Levi--Civita connection with ansatz of
type (\ref{genans3}) is subjected additionally to the conditions (\ref%
{frconstr1}), in this case on fractional function $\mathbf{\ }^{\alpha
}\varsigma _{\Upsilon }.$ For instance, we can chose that $\ \ \mathbf{\ }%
_{2}^{\alpha }n_{k}\left( x^{i}\right) =0$ and $\mathbf{\ }_{1}^{\alpha
}n_{k}\left( x^{i}\right) $ are any functions satisfying the conditions $\
\overset{\alpha }{\underline{\partial }}_{i}\ \ \mathbf{\ }_{1}^{\alpha
}n_{k}=\overset{\alpha }{\underline{\partial }}_{k}\ \ \mathbf{\ }%
_{1}^{\alpha }n_{i}.$ The constraints on values $\mathbf{\ }^{\alpha }w_{i}=-%
\overset{\alpha }{\underline{\partial }}_{i}\mathbf{\ }^{\alpha }\varsigma
_{\Upsilon }/\mathbf{\ }^{\alpha }\varsigma _{\Upsilon }^{\ast }$ result in
constraints on $\mathbf{\ }^{\alpha }\varsigma _{\Upsilon },$ which is
determined by $\mathbf{\ }^{\alpha }\Upsilon _{2}$ and $\mathbf{\ }^{\alpha
}f,$
\begin{eqnarray}
\left( \mathbf{\ }^{\alpha }w_{i}[\mathbf{\ }^{\alpha }\varsigma _{\Upsilon
}]\right) ^{\ast }+\mathbf{\ }^{\alpha }w_{i}[\mathbf{\ }^{\alpha }\varsigma
_{\Upsilon }]\left( \mathbf{\ }^{\alpha }h_{4}[\mathbf{\ }^{\alpha
}\varsigma _{\Upsilon }]\right) ^{\ast }+\overset{\alpha }{\underline{%
\partial }}_{i}\mathbf{\ }^{\alpha }h_{4}[\mathbf{\ }^{\alpha }\varsigma
_{\Upsilon }]=0, &&  \notag \\
\overset{\alpha }{\underline{\partial }}_{i}\mathbf{\ }^{\alpha }w_{k}[%
\mathbf{\ }^{\alpha }\varsigma _{\Upsilon }]=\overset{\alpha }{\underline{%
\partial }}_{k}\mathbf{\ }^{\alpha }w_{i}[\mathbf{\ }^{\alpha }\varsigma
_{\Upsilon }], &&  \label{auxc3}
\end{eqnarray}%
where, for instance, we denoted by $\mathbf{\ }^{\alpha }h_{4}[\mathbf{\ }%
^{\alpha }\varsigma _{\Upsilon }]$ the functional dependence on $\mathbf{\ }%
^{\alpha }\varsigma _{\Upsilon }.$ Such conditions are always satisfied for
cosmological solutions with $\mathbf{\ }^{\alpha }f=\mathbf{\ }^{\alpha
}f(v).$ For $\mathbf{\ }^{\alpha }\widehat{\mathbf{D}},$ if $\ \mathbf{\ }%
^{\alpha }\Upsilon _{2}=0$ and $\mathbf{\ }^{\alpha }\phi =const,$ the
coefficients $\mathbf{\ }^{\alpha }w_{i}(x^{k},v)$ can be arbitrary
functions. The simplest parametrization are for $\mathbf{\ }^{\alpha
}\varsigma _{\Upsilon }=1;$ this does not impose a functional dependence of $%
\mathbf{\ }^{\alpha }w_{i}$ on $\mathbf{\ }^{\alpha }\varsigma _{\Upsilon })$
as follows from (\ref{4ep3a}) with zero $\mathbf{\ }^{\alpha }\beta $ and $%
\mathbf{\ }^{\alpha }\alpha _{i},$ see (\ref{auxphi}). To generate solutions
for $\mathbf{\ }^{\alpha }\nabla $ such $\mathbf{\ }^{\alpha }w_{i}$ must be
additionally constrained following formulas (\ref{auxc3}) re--written for $%
\mathbf{\ }^{\alpha }w_{i}[\mathbf{\ }^{\alpha }\varsigma _{\Upsilon
}]\rightarrow \mathbf{\ }^{\alpha }w_{i}(x^{k},v)$ and $\mathbf{\ }^{\alpha
}h_{4}[\mathbf{\ }^{\alpha }\varsigma _{\Upsilon }]\rightarrow \mathbf{\ }%
^{\alpha }h_{4}\left( x^{i},v\right) .$

Finally, we emphasize that any solution $\mathbf{\ }^{\alpha }\mathbf{g}=\{%
\mathbf{\ }^{\alpha }g_{\alpha ^{\prime }\beta ^{\prime }}(u^{\alpha
^{\prime }})\}$ of the Einstein equations (\ref{cdeinst}) and/or (\ref%
{einsteq}) with Killing symmetry $\overset{\alpha }{\underline{\partial }}%
_{y}$ (for local coordinates in the form $y^{3}=v$ and $y^{4}=y)$ can be
parametrized in a form derived in this section. Using frame transforms of
type $\mathbf{\ }^{\alpha }e_{\alpha }=e_{\ \alpha }^{\alpha ^{\prime }}%
\mathbf{\ }^{\alpha }e_{\alpha ^{\prime }},$ with $\mathbf{\ }^{\alpha }%
\mathbf{g}_{\alpha \beta }=e_{\ \alpha }^{\alpha ^{\prime }}e_{\ \beta
}^{\beta ^{\prime }}\mathbf{\ }^{\alpha }g_{\alpha ^{\prime }\beta ^{\prime
}},$ for any $\mathbf{\ }^{\alpha }\mathbf{g}_{\alpha \beta }$ (\ref{gsol}),
we relate the class of such solutions, for instance, to the family of
metrics of type (\ref{genans}).

\section{Fractional Spacetimes and Black Holes}

\label{sbhfs} A fractional spacetime is with very different rules of computing local
partial derivatives  (via integration) of type $_{\ _{1}x^{i}}\overset{%
\alpha }{\underline{\partial }}_{i}$ (\ref{lfcd}) instead of usual partial
derivatives $\partial _{i}.$ The actions of such operators are very
different on singular functions, for instance, parametrized in the form (\ref%
{sfunct}). So, the singular etc structure of solutions of fractional
Einstein equations (\ref{cdeinst}) and/or (\ref{einsteq}) should differ substantially
from that for integer dimensions in general relativity.

There is a series of very
important questions to be solved in fractional models of gravity: 1) for
instance, if such theories contain black holes solutions? \ 2) if such
solutions for fractional black holes can be constructed, what are their
properties? 3) what may happen with an integer dimensional black hole into a
fractional background, for instance, of solitons? We should also analyze the problems:  4)
what may happen with integer and non--integer dimensional black holes under
fractional/nonholonomic Ricci flows? 5) what is the status of singular
solutions and horizons in fractional classical and quantum gravity etc.

In this section, we prove that black holes really exist  in fractional gravity
contrary to the hope that involving a new type of derivative calculus (\ref%
{lfcd}), and changing respectively the differential spacetime structure, we
may eliminate ''ambiguities'' with singularities etc. The concepts of black
hole, singularity and horizon seem to be fundamental ones for
various types of holonomic and nonholonomic, commutative and noncommutative,
pseudo--Riemanann and Finlser like, fractional and integer etc theories of
gravity. We also provide solutions for the questions 1--3) above and leave
4) and 5) for our future investigations.

\subsection{Fractional deformations of the Schwarzschild spacetime}

We consider a diagonal integer dimensional metric $~^{\varepsilon }\mathbf{g}
$ depending on a small real parameter $1>\varepsilon \gtrsim 0,$
\begin{equation}
~^{\varepsilon }\mathbf{g}=-d\xi \otimes d\xi -r^{2}(\xi )\ d\vartheta
\otimes d\vartheta -r^{2}(\xi )\sin ^{2}\vartheta \ d\varphi \otimes
d\varphi +\varpi ^{2}(\xi )\ dt\otimes \ dt.  \label{5aux1}
\end{equation}%
The local coordinates and nontrivial metric coefficients are parametriz\-ed:
\begin{eqnarray}
x^{1} &=&\xi ,x^{2}=\vartheta ,y^{3}=v=\varphi ,y^{4}=t,  \label{5aux1a} \\
\check{g}_{1} &=&-1,\ \check{g}_{2}=-r^{2}(\xi ),\ \check{h}_{3}=-r^{2}(\xi
)\sin ^{2}\vartheta ,\ \check{h}_{4}=\varpi ^{2}(\xi ),  \notag \\
\mbox{ \ for \ }\xi &=&\int dr\ \left| 1-\frac{2\mu _{0}}{r}+\frac{%
\varepsilon }{r^{2}}\right| ^{1/2}\mbox{\ and\ }\varpi ^{2}(r)=1-\frac{2\mu
_{0}}{r}+\frac{\varepsilon }{r^{2}}.  \notag
\end{eqnarray}%
For $\varepsilon =0$ in variable $\xi (r)$ and coefficients, the metric (\ref%
{5aux1}) is just the the Schwarzschild solution written in spacetime
spherical coordinates $(r,\vartheta ,\varphi ,t)$ with a point mass $\mu
_{0}.$

We search for a class of exact fractional vacuum solutions of type (\ref%
{gsol}) when the fractional metrics are generated by nonholonomic
deformations $\mathbf{\ }^{\alpha }g_{i}=\mathbf{\ }^{\alpha }\eta _{i}%
\check{g}_{i}$ and $h_{a}=\mathbf{\ }^{\alpha }\eta _{a}\check{h}_{a}$ and
some nontrivial $\mathbf{\ }^{\alpha }w_{i},\mathbf{\ }^{\alpha }n_{i},$
where $(\check{g}_{i},\check{h}_{a})$ are given by data (\ref{5aux1a}) and
parametrized by ansatz
\begin{eqnarray}
~_{\eta }^{\varepsilon }\mathbf{g} &=&-\mathbf{\ }^{\alpha }\eta _{1}(\xi
,\vartheta ,\theta )\mathbf{\ }^{\alpha }d\xi \otimes \mathbf{\ }^{\alpha
}d\xi -\eta _{2}(\xi ,\vartheta ,\theta )r^{2}(\xi )\ \mathbf{\ }^{\alpha
}d\vartheta \otimes \mathbf{\ }^{\alpha }d\vartheta  \label{5sol1} \\
&&-\mathbf{\ }^{\alpha }\eta _{3}(\xi ,\vartheta ,\varphi ,\theta )r^{2}(\xi
)\sin ^{2}\vartheta \ \mathbf{\ }^{\alpha }\delta \varphi \otimes \mathbf{\ }%
^{\alpha }\delta \varphi  \notag \\
&&+\mathbf{\ }^{\alpha }\eta _{4}(\xi ,\vartheta ,\varphi ,\theta )\varpi
^{2}(\xi )\ \mathbf{\ }^{\alpha }\delta t\otimes \mathbf{\ }^{\alpha }\delta
t,  \notag \\
\mathbf{\ }^{\alpha }\delta \varphi &=&\mathbf{\ }^{\alpha }d\varphi +%
\mathbf{\ }^{\alpha }w_{1}(\xi ,\vartheta ,\varphi ,\theta )\mathbf{\ }%
^{\alpha }d\xi +\mathbf{\ }^{\alpha }w_{2}(\xi ,\vartheta ,\varphi ,\theta )%
\mathbf{\ }^{\alpha }d\vartheta ,\   \notag \\
\mathbf{\ }^{\alpha }\delta t &=&\mathbf{\ }^{\alpha }dt+\mathbf{\ }^{\alpha
}n_{1}(\xi ,\vartheta ,\theta )\mathbf{\ }^{\alpha }d\xi +\mathbf{\ }%
^{\alpha }n_{2}(\xi ,\vartheta ,\theta )\mathbf{\ }^{\alpha }d\vartheta ,
\notag
\end{eqnarray}%
when the coefficients will be constructed determine solutions of the system
of equations (\ref{4ep1a})--(\ref{4ep4a}) with $\mathbf{\ }^{\alpha
}\Upsilon _{\beta }=0.$

The equation (\ref{4ep2a}) for $\mathbf{\ }^{\alpha }\Upsilon _{2}=0$ is
solved by any%
\begin{eqnarray}
\mathbf{\ }^{\alpha }h_{3} &=&-\mathbf{\ }_{0}^{\alpha }h^{2}(\mathbf{\ }%
^{\alpha }b^{\ast })^{2}=\mathbf{\ }^{\alpha }\eta _{3}(\xi ,\vartheta
,\varphi ,\theta )r^{2}(\xi )\sin ^{2}\vartheta ,  \label{aux41} \\
\mathbf{\ }^{\alpha }h_{4} &=&\mathbf{\ }^{\alpha }b^{2}=\mathbf{\ }^{\alpha
}\eta _{4}(\xi ,\vartheta ,\varphi ,\theta )\varpi ^{2}(\xi ),  \notag
\end{eqnarray}%
for $|\mathbf{\ }^{\alpha }\eta _{3}|=(\mathbf{\ }_{0}^{\alpha }h)^{2}|%
\check{h}_{4}/\check{h}_{3}|\left[ \left( \sqrt{|\mathbf{\ }^{\alpha }\eta
_{4}|}\right) ^{\ast }\right] ^{2}.$ We consider $\mathbf{\ }_{0}^{\alpha
}h=const$ (it must be $\mathbf{\ }_{0}^{\alpha }h=2$ in order to satisfy the
condition (\ref{4ep2a}) with zero source), where $\mathbf{\ }^{\alpha }\eta
_{4}$ can be any function satisfying the condition $\mathbf{\ }^{\alpha
}\eta _{4}^{\ast }\neq 0.$ This way, it is possible to generate a class of
solutions for any function $\mathbf{\ }^{\alpha }b(\xi ,\vartheta ,\varphi
,\theta )$ with $\mathbf{\ }^{\alpha }b^{\ast }\neq 0.$ For classes of
solutions with nontrivial sources, it is more convenient to work directly
with fractional polarizations $\mathbf{\ }^{\alpha }\eta _{4},$ for $\mathbf{%
\ }^{\alpha }\eta _{4}^{\ast }\neq 0.$ In another turn, for vacuum
configurations, it is better to chose as a generating function, for
instance, $\mathbf{\ }^{\alpha }h_{4},$ for $\mathbf{\ }^{\alpha
}h_{4}^{\ast }\neq 0.$ The fractional polarizations $\mathbf{\ }^{\alpha
}\eta _{1}$ and $\mathbf{\ }^{\alpha }\eta _{2},$ when $\mathbf{\ }^{\alpha
}\eta _{1}=\mathbf{\ }^{\alpha }\eta _{2}r^{2}=e\mathbf{\ }^{^{\mathbf{\ }%
^{\alpha }}\psi (\xi ,\vartheta )}$ (for fractional configurations, this is
not just the exponential function but its fractional version), from (\ref%
{4ep1a}) with $\mathbf{\ }^{\alpha }\Upsilon _{4}=0,$ i.e. from $\mathbf{\ }%
^{\alpha }\psi ^{\bullet \bullet }+\mathbf{\ }^{\alpha }\psi ^{\prime \prime
}=0.$

Putting the above coefficient in (\ref{5sol1}), we construct a class of
exact vacuum solutions in fractional gravity defining stationary fractional
nonholonomic deformations on a small parameter $\varepsilon $ of the
Sch\-warz\-schild metric,
\begin{eqnarray}
~_{\varepsilon }^{\alpha }\mathbf{g} &=&-e^{\mathbf{\ }^{\alpha }\psi (\xi
,\vartheta ,\theta )}\left( \mathbf{\ }^{\alpha }d\xi \otimes \mathbf{\ }%
^{\alpha }d\xi +\ \mathbf{\ }^{\alpha }d\vartheta \otimes \mathbf{\ }%
^{\alpha }d\vartheta \right)  \label{5sol1a} \\
&&-4\left[ \left( \sqrt{|\mathbf{\ }^{\alpha }\eta _{4}(\xi ,\vartheta
,\varphi ,\theta )|}\right) ^{\ast }\right] ^{2}\varpi ^{2}(\xi )\mathbf{\ }%
^{\alpha }\ \delta \varphi \otimes \ \mathbf{\ }^{\alpha }\delta \varphi
\notag \\
&&+\mathbf{\ }^{\alpha }\eta _{4}(\xi ,\vartheta ,\varphi ,\theta )\varpi
^{2}(\xi )\ \mathbf{\ }^{\alpha }\delta t\otimes \mathbf{\ }^{\alpha }\delta
t,  \notag \\
\mathbf{\ }^{\alpha }\delta \varphi &=&\mathbf{\ }^{\alpha }d\varphi +%
\mathbf{\ }^{\alpha }w_{1}(\xi ,\vartheta ,\varphi ,\theta )\mathbf{\ }%
^{\alpha }d\xi +\mathbf{\ }^{\alpha }w_{2}(\xi ,\vartheta ,\varphi ,\theta )%
\mathbf{\ }^{\alpha }d\vartheta ,\   \notag \\
\mathbf{\ }^{\alpha }\delta t &=&\mathbf{\ }^{\alpha }dt+\ \mathbf{\ }%
_{1}^{\alpha }n_{1}(\xi ,\vartheta ,\theta )\mathbf{\ }^{\alpha }d\xi +\
\mathbf{\ }_{2}^{\alpha }n_{2}(\xi ,\vartheta ,\theta )\mathbf{\ }^{\alpha
}d\vartheta .  \notag
\end{eqnarray}

In general, the solutions for fractional metrics (\ref{5sol1}), or (\ref%
{5sol1a}), do not define black holes and do not describe obvious physical
situations. We can consider that for general nonholonomic fractional
deformations a usual ''integer'' black hole may ''dissipate'' into a
fractional structure of a more sophisticate fractional spacetime ether with
a not defined status of singularities of coefficients of metric.

In next subsections, we show that it is possible to chose certain
nonholonomic distributions when the singular character of the coefficient $%
\varpi ^{2}(\xi )$ vanishing on the horizon of a Schwarzschild black hole
result in physical properties of usual black holes.

\subsection{Non--integer gravitational ellipsoid configurations}

It is possible to extract a class of metrics (\ref{5sol1a}) defining
fractional deformations of the Schwarzschild solution depending on parameter
$\varepsilon $ with possible physical interpretation of fractional
gravitational vacuum configurations with spherical and/or rotoid (ellipsoid)
symmetry. We chose in (\ref{aux41}) \ a generating functions of type
\begin{equation}
\mathbf{\ }^{\alpha }b^{2}=\mathbf{\ }^{\alpha }q(\xi ,\vartheta ,\varphi
)+\varepsilon \mathbf{\ }^{\alpha }s(\xi ,\vartheta ,\varphi )  \label{gf1}
\end{equation}%
considering, for simplicity, only linear decompositions on a small parameter
$\varepsilon .$ For (\ref{gf1}), we get
\begin{equation*}
\left( \mathbf{\ }^{\alpha }b^{\ast }\right) ^{2}=\left[ (\sqrt{|\mathbf{\ }%
^{\alpha }q|})^{\ast }\right] ^{2}\left[ 1+\varepsilon \left( \mathbf{\ }%
^{\alpha }s/\sqrt{|\mathbf{\ }^{\alpha }q|}\right) ^{\ast }/(\sqrt{|\mathbf{%
\ }^{\alpha }q|})^{\ast }\right] .
\end{equation*}%
So, we can compute in (\ref{5sol1a}) the coefficients $\mathbf{\ }^{\alpha
}h_{3}$ and $\mathbf{\ }^{\alpha }h_{4}$ and corresponding polarizations $%
\mathbf{\ }^{\alpha }\eta _{3}$ and $\mathbf{\ }^{\alpha }\eta _{4},$ using
formulas (\ref{aux41}). Here we note that if we put $\varepsilon =0$ we can
generate fractional deformations of the Schwarzschild solution not depending
on parameter $\alpha ,$ when $\mathbf{\ }^{\alpha }b^{2}=\mathbf{\ }^{\alpha
}q$ and $\left( \mathbf{\ }^{\alpha }b^{\ast }\right) ^{2}=\left[ (\sqrt{|%
\mathbf{\ }^{\alpha }q|})^{\ast }\right] ^{2}.$

Fractional rotoid configurations are generated for
\begin{equation}
\mathbf{\ }^{\alpha }q=1-\frac{2\mathbf{\ }^{\alpha }\mu (\xi ,\vartheta
,\varphi )}{r}\mbox{ and }\mathbf{\ }^{\alpha }s=\frac{q_{0}(r)}{4\mathbf{\ }%
^{\alpha }\mu ^{2}}\sin (\omega _{0}\varphi +\varphi _{0}),  \label{aux42}
\end{equation}%
when $\mathbf{\ }^{\alpha }\mu (\xi ,\vartheta ,\varphi )=\mu _{0}+\mu
_{1}(\xi ,\vartheta ,\varphi )$ \ describes fractional locally anisotropic
polarized mass. The constants $\mu _{0},\omega _{0}$ and $\varphi _{0}$ and
arbitrary functions\newline
$\mu _{1}(\xi ,\vartheta ,\varphi )$ and $q_{0}(r)$ have to be defined from
some boundary conditions, with $\ \varepsilon $ treated as the eccentricity%
\footnote{$\varepsilon $ is an eccentricity because, for integer
configurations, the coefficient $h_{4}=b^{2}=\eta _{4}(\xi ,\vartheta
,\varphi ,\ \bar{\theta})$ $\varpi ^{2}(\xi )$ becomes zero for data (\ref%
{aux42}) if $r_{+}\simeq {2\mu _{0}}/[{1+\varepsilon \frac{q_{0}(r)}{4\mu
^{2}}\sin (\omega _{0}\varphi +\varphi _{0})}];$ this is the ''parametric''
equation for an ellipse $r_{+}(\varphi )$ for any fixed values $\frac{%
q_{0}(r)}{4\mu ^{2}},\omega _{0},\varphi _{0}$ and $\mu _{0}$} of an
ellipsoid.

The fractional stationary rotoid solutions for the Schwarzschild metric in
general relativity can be written in the form
\begin{eqnarray}
~_{rot\ \varepsilon }^{\alpha }\mathbf{g} &=&-e^{\ ^{\alpha }\psi }\left(
\mathbf{\ }^{\alpha }d\xi \otimes \mathbf{\ }^{\alpha }d\xi +\ \mathbf{\ }%
^{\alpha }d\vartheta \otimes \mathbf{\ }^{\alpha }d\vartheta \right)  \notag
\\
&&-4\ \left[ (\sqrt{|\mathbf{\ }^{\alpha }q|})^{\ast }\right] ^{2}\left[
1+\varepsilon \left( \mathbf{\ }^{\alpha }s/\sqrt{|\mathbf{\ }^{\alpha }q|}%
\right) ^{\ast }/(\sqrt{|\mathbf{\ }^{\alpha }q|})^{\ast }\right] \mathbf{\ }%
^{\alpha }\delta \varphi \otimes \ \mathbf{\ }^{\alpha }\delta \varphi
\notag \\
&&+\left( \mathbf{\ }^{\alpha }q+\varepsilon \mathbf{\ }^{\alpha }s\right) \
\mathbf{\ }^{\alpha }\delta t\otimes \mathbf{\ }^{\alpha }\delta t,
\label{rotoidm} \\
\mathbf{\ }^{\alpha }\delta \varphi &=&\mathbf{\ }^{\alpha }d\varphi +%
\mathbf{\ }^{\alpha }w_{1}\mathbf{\ }^{\alpha }d\xi +\mathbf{\ }^{\alpha
}w_{2}\mathbf{\ }^{\alpha }d\vartheta ,  \notag \\
\mathbf{\ }^{\alpha }\delta t &=&\mathbf{\ }^{\alpha }dt+\ \mathbf{\ }%
_{1}^{\alpha }n_{1}\mathbf{\ }^{\alpha }d\xi +\ \mathbf{\ }_{1}^{\alpha
}n_{2}\mathbf{\ }^{\alpha }d\vartheta ,  \notag
\end{eqnarray}%
where functions $\mathbf{\ }^{\alpha }q(\xi ,\vartheta ,\varphi )$ and $%
\mathbf{\ }^{\alpha }s(\xi ,\vartheta ,\varphi )$ are given by formulas (\ref%
{aux42}). The N--connec\-ti\-on coefficients, $\mathbf{\ }^{\alpha
}w_{i}(\xi ,\vartheta ,\varphi )$ and $\ \mathbf{\ }^{\alpha }n_{i}=$ $\ \
\mathbf{\ }_{1}^{\alpha }n_{i}(\xi ,\vartheta ),$ are subjected to
conditions of type (\ref{frconstr1}),
\begin{eqnarray*}
\mathbf{\ }^{\alpha }w_{1}\mathbf{\ }^{\alpha }w_{2}\left( \ln |\mathbf{\ }%
^{\alpha }w_{1}/\mathbf{\ }^{\alpha }w_{2}|\right) ^{\ast } &=&\mathbf{\ }%
^{\alpha }w_{2}^{\bullet }-\mathbf{\ }^{\alpha }w_{1}^{\prime },\quad
\mathbf{\ }^{\alpha }w_{i}^{\ast }\neq 0; \\
\mbox{ or \ }\mathbf{\ }^{\alpha }w_{2}^{\bullet }-\mathbf{\ }^{\alpha
}w_{1}^{\prime } &=&0,\quad \mathbf{\ }^{\alpha }w_{i}^{\ast }=0;\  \\
\ \mathbf{\ }_{1}^{\alpha }n_{1}^{\prime }(\xi ,\vartheta )-\ \ \mathbf{\ }%
_{1}^{\alpha }n_{2}^{\bullet }(\xi ,\vartheta ) &=&0
\end{eqnarray*}%
and $\mathbf{\ }^{\alpha }\psi (\xi ,\vartheta )$ being any function for
which $\mathbf{\ }^{\alpha }\psi ^{\bullet \bullet }+\mathbf{\ }^{\alpha
}\psi ^{\prime \prime }=0.$

For small eccentricities, a metric (\ref{rotoidm}) defines stationary
fractional configurations for so--called black ellipsoid type solutions.
Their stability and properties can be stated and analyzed by adapting to
fractional calculus the methods elaborated in \cite{vbe1,vbe2,vncg} (a
summary of results and generalizations for various types of locally
anisotropic gravity models in Ref. \cite{vsgg}, see similar constructions
for a noncommutative Finsler version of rotoid spacetimes in \cite{vcg}).

\subsection{Stationary black holes in fractional solitonic configurations}

Finally, we analyze two types of fractional rotoid solutions (in particular
the Schwarzschild one) imbedded into 1) integer solitonic background and 2)
into a fractional solitonic background defined by an exact solution of a
fractional differential equation.

\subsubsection{Integer solitonic background}

There are static three dimensional solitonic distributions $\eta (\xi
,\vartheta ,\varphi ,\theta ),$ defined as solutions of an (integer)
solitonic equation
\begin{equation*}
\eta ^{\bullet \bullet }+\epsilon (\eta ^{\prime }+6\eta \ \eta ^{\ast
}+\eta ^{\ast \ast \ast })^{\ast }=0,\ \epsilon =\pm 1,
\end{equation*}%
resulting in stationary black ellipsoid--solitonic fractional metrics \
generated as further deformations of a metric $~~_{rot\ \varepsilon
}^{\alpha }\mathbf{g}$ (\ref{rotoidm}).\footnote{%
a function $\eta $ can be a solution of any three dimensional solitonic and/
or other nonlinear wave equations} The generated fractional solitonic
gravitational metrics are of type
\begin{eqnarray}
\ ^{\alpha }\mathbf{g} &=&-e^{\ ^{\alpha }\psi }\left( \ ^{\alpha }d\xi
\otimes \ ^{\alpha }d\xi +\ \ ^{\alpha }d\vartheta \otimes \ ^{\alpha
}d\vartheta \right) -  \label{solrot} \\
&&4\left[ (\sqrt{|\eta \ ^{\alpha }q|})^{\ast }\right] ^{2}\left[
1+\varepsilon \frac{1}{(\sqrt{|\eta \ ^{\alpha }q|})^{\ast }}\left( \frac{\
^{\alpha }s}{\sqrt{|\eta \ ^{\alpha }q|}}\right) ^{\ast }\right] \ ^{\alpha
}\delta \varphi \otimes \ ^{\alpha }\delta \varphi  \notag \\
&&+\eta \left( \ ^{\alpha }q+\varepsilon \ ^{\alpha }s\right) \ ^{\alpha
}\delta t\otimes \ ^{\alpha }\delta t,  \notag \\
\ ^{\alpha }\delta \varphi &=&\ ^{\alpha }d\varphi +\ ^{\alpha }w_{1}\
^{\alpha }d\xi +\ ^{\alpha }w_{2}\ ^{\alpha }d\vartheta ,  \notag \\
\ ^{\alpha }\delta t &=&\ ^{\alpha }dt+\ \ _{1}^{\alpha }n_{1}\ ^{\alpha
}d\xi +\ \ _{1}^{\alpha }n_{2}\ ^{\alpha }d\vartheta ,  \notag
\end{eqnarray}%
where the N--connection coefficients are taken the same as for (\ref{rotoidm}%
).

The metrics (\ref{solrot}) are of type (\ref{5sol1a}). So, they positively
define vacuum solutions of fractional Einstein equations.

\subsubsection{Fractional solitonic backgrounds}

An interesting property of the solutions (\ref{genans3}) is that they depend
on a generating fractional function $\mathbf{\ }^{\alpha }f^{2}\left(
x^{i},v\right) $ which is a general one, but constrained to some conditions
in order to generate a corresponding class of solutions. For instance, we
can consider that the fractional metric (\ref{solrot}) is additionally
deformed nonholonomically by a solution $\ ^{\alpha }\rho (v),$ when $\
^{\alpha }\rho ^{\ast }=\overset{\alpha }{\underline{\partial }}_{v}\rho
=_{\ _{1}v}\overset{\alpha }{\underline{\partial }}_{v}{}\ ^{\alpha }\rho $
and $\ \ _{\ _{1}v}\overset{\alpha }{\partial }_{v}$ being the left RL
derivative on $v,$ of a fractional differential equation%
\begin{equation}
\ \ \ _{\ _{1}x}\overset{\alpha }{\partial }_{x}(\mathbf{\ }^{\alpha }\rho
^{\ast })+\ ^{1}z(v)\mathbf{\ }^{\alpha }\rho ^{\ast }=\ ^{2}z(v)
\label{auxf}
\end{equation}%
for some suitable functions $\ ^{1}z(v)\mathbf{\ }$\ and $\ ^{2}z(v).\mathbf{%
\ }$This fractional equation can be solved in general form \cite{baltr},%
\begin{equation}
\ ^{\alpha }\rho (v)=\sum\limits_{p=0}^{\infty }(-1)^{p}\ _{\ _{1}v}\overset{%
\alpha }{I}_{v}\left[ \ ^{1}z(v)^{-1}\ _{\ _{1}v}\overset{\alpha }{\partial }%
_{v}\right] ^{p}\left\{ \frac{\ ^{2}z(v)}{\ ^{1}z(v)}\right\} +\ ^{1}c(v-\
^{1}v)+\ ^{2}c,  \label{auxfs}
\end{equation}%
where $\ ^{1}c$ and $\ ^{2}c$ are constants. If $\ ^{1}z(v)=0$ and $\
^{2}z(v)=\lambda \ ^{\alpha }\rho (v),$ where $\lambda \in \mathbb{R},$ the
equation (\ref{auxf}) transform into a fractional Euler--Lagrange equation
for the one--dimensional fractional oscillator. Its solution is different
from (\ref{auxfs}) but also can be expressed as a series (see details in %
\cite{baltr}).

We construct a fractional black rotoid solution embedded both into an
integer solitonic background and a one--dimensional fractional gravitational
generalized oscillator (\ref{auxfs}) if we consider the (vacuum solution)
metric
\begin{eqnarray*}
\ ^{\alpha }\mathbf{g} &=&-e^{\ ^{\alpha }\psi }\left( \ ^{\alpha }d\xi
\otimes \ ^{\alpha }d\xi +\ \ ^{\alpha }d\vartheta \otimes \ ^{\alpha
}d\vartheta \right) -4\left[ (\sqrt{|\eta \ ^{\alpha }q\ ^{\alpha }\rho |}%
)^{\ast }\right] ^{2}\times \\
&&\left[ 1+\varepsilon \frac{1}{(\sqrt{|\eta \ ^{\alpha }q\ ^{\alpha }\rho |}%
)^{\ast }}\left( \frac{\ ^{\alpha }s}{\sqrt{|\eta \ ^{\alpha }q\ ^{\alpha
}\rho |}}\right) ^{\ast }\right] \ ^{\alpha }\delta \varphi \otimes \
^{\alpha }\delta \varphi \\
&&+\eta \ ^{\alpha }\rho \left( \ ^{\alpha }q+\varepsilon \ ^{\alpha
}s\right) \ ^{\alpha }\delta t\otimes \ ^{\alpha }\delta t, \\
\ ^{\alpha }\delta \varphi &=&\ ^{\alpha }d\varphi +\ ^{\alpha }w_{1}\
^{\alpha }d\xi +\ ^{\alpha }w_{2}\ ^{\alpha }d\vartheta , \\
\ ^{\alpha }\delta t &=&\ ^{\alpha }dt+\ \ _{1}^{\alpha }n_{1}\ ^{\alpha
}d\xi +\ \ _{1}^{\alpha }n_{2}\ ^{\alpha }d\vartheta ,
\end{eqnarray*}%
where the N--connection coefficients are taken the same as for (\ref{rotoidm}%
). This type of metrics are also of type (\ref{5sol1a}), with a different
nonholonomic structure, and also define vacuum solutions of fractional
Einstein equations.

The fractional nonholonomic black hole type solutions (\ref{rotoidm}) and (%
\ref{solrot}), and related solitonic and fractional oscillator extensions,
are stationary.

\section{Discussion, Conclusions and Further Developments}

\label{sconcl}

Different concepts of dimension and related definitions of derivatives
(Riemann--Lesbegue, Caputo etc) and integrals (Lesbegue--Stieltjes,
Riemann--Stieltjes, fractional integrals etc) were introduced long time ago
and studied at present in various branches of science. In this work, we
developed a geometric approach to fractional gravity theories based on
fractional nonholonomic manifolds, nonlinear connections and generalized
solutions of fractional Einstein equations, deriving all constructions from
the Caputo left fractional derivative. Following such a direction, we
preserve a maximally possible analogy between spaces of integer and
non--integer dimensions. For study of general dynamical and evolution
theories of gravitational and matter fields (classical and quantum models),
we have to apply various methods from the geometry of nonholonomic manifolds
(originally elaborated for Lagrange--Finsler spaces) developed for a
correspondingly adapted fractional calculus.

In our approach, the fractional gravity can be naturally defined on
fractional manifolds following the same principles as in general relativity.
More than that, there is an unified method allowing to derive both types of
metric compatible generalizations of the Einstein theories, with a
fractional version of the Levi--Civita connection, and of generalized
Lagrange--Finsler models, with the canonical/Cartan distinguished
connections. The surprising result is that the fractional Einstein equations
can be integrated in general form by adapting the constructions with respect
to certain classes of frame elongated linearly by a nonlinear connection
(N--connection) structure. Both for the fractional and integer dimension
(pseudo) Riemannian spaces, the coefficients N--connections are defined by
certain off--diagonal terms of metrics. For Lagrange--Finsler theories, the
N--connections are determined as a fundamental geometric structure induced
by a fractional/integer generating function.

In this work, we classified possible integral varieties of fractional
nonholonomic gravitational field equations. We also provided a study of
Schwarz\-schild type black holes and rotoid configurations in fractional
spacetimes. The main conclusion was that the black holes exist also in
fractional gravity and that such configurations may survive for certain
types of small nonholonomic fractional deformations of metrics and in
integer, or fractional, solitonic backgrounds. Nevertheless, more general
classes of fractional nonholonomic transforms result in exact solutions with
not obvious physical interpretation.

There are many directions of investigation left to explore in fractional
calculus and geometry and applications in various sciences. Here we outline
seven important ones being related to possible developments of our former
results in (integer dimensional) geometry of nonholonomic manifolds/
bundles, and applications to modern classical and quantum physics theories:

\begin{enumerate}
\item \textit{Foundations of fractional classical and quantum field theories
}with possible noncommutative, supersymmetric, string, brane generalizations
etc. There is a number of publications in such directions \cite%
{bal06,balmus05,calcagni1,calcagni2,carp,herrmann,lask2,
musln,taras06,taras08,zasl2,zavada}. Our proposal is to analyze if such
constructions can be redefined for nonholonomic fractional noncommutative/
supersymmetric etc spaces and string/brane models, see reviews of such
results in our works \cite{vstr1,vstr2,vmon1,vmon2,vsgg}, when for adapted
nonholonomic distributions the constructions the fundamental field equations
became integrable in very general forms.

\item \textit{A theory of fractional nonholonomic Ricci flows } was proposed
recently in \cite{vfrrf}. This generalize in a fractional fashion a series
of our papers on nonholonomic Ricci flows of Einstein, Lagrange--Finsler and
noncommutative/ nonsymmetric spaces \cite{vnhrf1,vnhrf1,vrfnc} and exact
solutions of the Einstein equations and Hamilton's equations and
applications in cosmology and astrophysics \cite{vnhrns,vnhrfs1,vnhrfs2}.
Solutions with fractional evolution of geometric objects present a strong
theoretical arguments for study such theories.

\item \textit{Exact solutions in fractional gravity and applications in
modern cosmology and astrophysics.} The first examples of such fractional
exact solutions were constructed and analyzed in previous two sections of
this paper. We proved that our version of fractional Einstein equations can
be solved in general form and analyzed models of black hole/ellipsoid
fractional metrics. Here we note that there are various attempts to relate
modern cosmological experimental data to effects of fractional/fractal
gravity and matter field interactions \cite%
{nab1,nab2,roberts,calcagni1,calcagni2}. With respect to our former works on
general/exact solutions in various models of gravity theories and
applications, we suppose to be of interest in modern mathematical physics
and gravity such issues: a) Exact solutions for models of fractional gravity
of arbitrary dimensions and with different types of connections generalizing %
\cite{vexsol,vsgg,ijgmmp}; b) locally anisotropic fractional cosmology
models \cite{vcosms,vsgg}; c) various fractional brane, noncommutative,
anisotropic Taub\ NUT, wormhole and black hole solutions \cite%
{vcg,vsingl1,vsingl2,vbe1,vbe2,vncg,vpopa,vts}.

\item \textit{Fractional Clifford spinor structures and Dirac operators }%
have been considered, for instance, in \cite{raspini2,mab}. Following an
''exactly integrable'' fractional nonholonomic approach, we find such
perspective for further developments on fractional Clifford geometry and
applications. There are at least five different sub--directions: a)
fractional Clifford--Lagrange/--Finsler/ - Hamilton spaces and higher order
generalizations, see original results for integer dimensional spaces in \cite%
{vsp1,vsp2,vvicol,vmon2}; b) fractional Clifford--Finsler algebroids,
extending \cite{valgebr}, and c) nonholonomic non--integer gerbes as
generalizations from \cite{vgerb}; d) exact solutions of fractional
Einstein--Dirac equations adapted to N--connection structure as we
considered in \cite{vpopa,vts} and e) fractional models of noncommutative
geometry, Dirac operators and noncommutative gravity and field interactions,
see original results in \cite{vrfnc,vncg,vvc} and Part III of \cite{vsgg}.

\item \textit{Fractional Lagrange--Finsler and Hamilton--Cartan spaces and
higher order generalizations. }This direction is originally considered in %
\cite{albu} as a fractional generalization (using the Riemann--Liouville
integral) of some results from \cite{ma,m1}. Here we cite our alternative/
complimentary  constructions with higher order (super) vector bundles/
Clifford bundles and super--strings \cite%
{vstr1,vsp2,vmon1,vmon2,vexsol,vden1,vden2} which can be generalized for
fractional higher order calculus, adapted to nonholonomic distributions and
fractional Caputo derivatives, with exactly integrable string/brane/ spinor
and modified gravitational field equations.

\item \textit{Fractional quantum gravity, deformation quantization,
renormalizability, and gauge models.} Such theories, for instance, in
fractional versions, are supposed to provide new tools in quantization of
gravity and matter field interactions, see \cite%
{calcagni1,lask1,lask2,taras05,taraspl}. \ To elaborate a geometric approach
to fractional quantum theories, including quantum gravity, is a topic for
future investigations. Working with nonholonomic distributions, such
constructions can be derived fractionally from our former results on
deformation quantization of gravity and Lagrange--Finsler and
Hamilton--Cartan spaces \cite{vdqlf,vdqla,avdqhc}, A--brane fractional
quantization \cite{vbrane} and bi--connection quantization \cite{vqgr1,vqgr2}

\item \textit{Diffusion, kinetics thermodynamics and fractional nonlinear
dynamics with solitons, chaos and fractals. }This is, perhaps, the most
elaborated direction in fractional calculus with a number of applications,
see \cite{calcagni2,carrer,zasl1,carp,taras05,zasl2,zasl3} and references
therein. Such fractional constructions adapted to nonholonomic distributions
are possible for locally anisotropic (in general, supersymmetric) stochastic
calculus, see \cite{vdif} and Chapters 5 and 10 in monograph \cite{vmon1},
and kinetic theories and anisotropic thermodynamics \cite{vkin1,vkin2}.
There is also recent our papers with bi--Hamilton structures, solitonic
hierarchies and encoding the solutions of gravitational field equations in
Einstein--Finsler gravity and Ricci--Lagrange theories \cite%
{ancov,vsolitgr,vsolrf}. Solitonic hierarchies encoding of solutions of fractional systems of equations related to fundamental theories present a substantial interest for fractional dynamics, evolution and field interactions.
\end{enumerate}

A series of our further works will be devoted to researches concerning above
directions.

\end{document}